\documentclass[useAMS,usenatbib]{mn2e} 
\usepackage{aas_macros}
\usepackage{graphics}
\usepackage[pdftex]{graphicx}
\usepackage{epstopdf}
\usepackage{epsfig}  
\usepackage{natbib} 
\usepackage{float}
\usepackage{amsmath}
\usepackage{times}
\usepackage[varg]{txfonts}
\usepackage{verbatim} 
\usepackage{multirow,bigdelim} 
\usepackage{color}
\usepackage{vmargin}
\setmarginsrb{1.2cm}{1cm}{1.2cm}{1cm}{1cm}{1cm}{1cm}{1cm}

  \makeatletter
    \renewcommand{\paragraph}{\@startsection{paragraph}{4}{\z@}%
      {-3.25ex\@plus -1ex \@minus -.2ex}%
      {1.5ex \@plus .2ex}%
      {\normalfont\small\centering}}
     
    \renewcommand{\subparagraph}{\@startsection{subparagraph}{5}{\z@}%
      {-3.25ex\@plus -1ex \@minus -.2ex}%
      {1.5ex \@plus .2ex}%
      {\normalfont\small\centering}}
    \makeatother

\newcommand{\hMpc}{{h$^{-1}$Mpc}}

\newcommand{\kms}{{ km~s$^{-1}$}}

\title[Minimization of Biases]{Minimization of Biases in Galaxy Peculiar Velocity Catalogs}
\author[Sorce]
{Jenny G. Sorce$^{1}$\thanks{E-mail: \texttt{jsorce@aip.de}}\\ 
$^1$Leibniz-Institut f\"{u}r Astrophysik, 14482 Potsdam, Germany\\
}

\begin{document}

\date{}

\pagerange{\pageref{firstpage}--\pageref{lastpage}} \pubyear{2015}

\maketitle

\label{firstpage}

\begin{abstract}

Galaxy distances and derived radial peculiar velocity catalogs constitute valuable datasets to study the dynamics of the Local Universe. However, such catalogs suffer from biases whose effects increase with the distance. Malmquist biases and lognormal error distribution affect the catalogs. Velocity fields of the Local Universe reconstructed with these catalogs present a spurious overall infall onto the Local Volume if they are not corrected for biases. Such an infall is observed in the reconstructed velocity field obtained when applying the Bayesian Wiener-Filter technique to the raw second radial peculiar velocity catalog of the Cosmicflows project. In this paper, an iterative method to reduce spurious non-Gaussianities in the radial peculiar velocity distribution, to retroactively derive overall better distance estimates resulting in a minimization of the effects of biases, is presented. This method is tested with mock catalogs. To control the cosmic variance, mocks are built out of different cosmological constrained simulations which resemble the Local Universe. To realistically reproduce the effects of biases, the mocks are constructed to be look-alikes of the second data release of the Cosmicflows project, with respect to the size, distribution of data and distribution of errors. Using a suite of mock catalogs, the outcome of the correction is verified to be affected neither by the added error realization, nor by the datapoint selection, nor by the constrained simulation. Results are similar for the different tested mocks. After correction, the general infall is satisfactorily suppressed. The method allows us to obtained catalogs which together with the Wiener-Filter technique give reconstructions approximating non biased velocity fields at 100-150\kms\ (2-3~\hMpc\ in terms of linear displacement), the linear theory threshold.
\end{abstract}

\begin{keywords}
methods: numerical, techniques: radial velocities - large-scale structure of universe
\end{keywords}

\section{Introduction}

Nodes, sheets, filaments and voids compose the Large Scale Structure of the Local Universe. Understanding the apparent complexity of gravitational motions that results in the formation of such an environment constitutes a real challenge. For instance, progress has been made over the past few years to better understand the motion of our Galaxy in this context \citep{2007ASPC..379...24T,2008glv..book....3T,2008ApJ...676..184T}.  Yet, in an era of "precision cosmology", only 30\% of the motion of the Milky Way relative to the Cosmic Microwave Background is understood exactly. Catalogs of accurate distances and radial peculiar velocities of galaxies constitute a valuable source of information to study our neighborhood. Peculiar velocities are direct tracers of the underlying gravitational fields as they account for the total amount of matter. This is an advantage over redshift surveys which, although they are more easily obtained, account solely for the luminous matter and thus are subject to biases \citep[e.g.][]{2013PASA...30...30B}. For that reason, Cosmicflows\footnote{http://www.ipnl.in2p3.fr/projet/cosmicflows/}, whose goal is to map the Local Universe, develops such radial peculiar velocity catalogs. The final aim is to reconstruct overdensity and velocity fields of our neighborhood using exclusively peculiar velocities which are highly linear, correlated on large scales and not affected by the luminous bias. A large amount of observational data is necessary to obtain accurate galaxy distances from which the radial peculiar velocities are derived. \citet{2013AJ....146...86T} released the second generation catalog, {\it cosmicflows-2}, of the project which reaches 150 \hMpc\ and contained more than 8,000 galaxy distance and peculiar velocity measurements. Although superior in size and extent to the first catalog of the project, this catalog is not, however, immune to biases when considered without special treatments as stipulated by Tully et al. 2013\footnote{"[...] although individual distances may be unbiased, the homogeneous and non-homogeneous Malmquist effects and error bias can generate spurious artifacts in velocity fields. These issues must be addressed if velocities are used to infer the distribution of matter." }. At such a depth, errors can be larger than peculiar velocities, overwhelming their signal. When reconstructing the full 3D velocity field, applying the Bayesian Wiener-Filter technique \citep[WF,][]{1995ApJ...449..446Z,1999ApJ...520..413Z} assuming by default a symmetric error distribution, the negative tail \citep[asymmetric with respect to the positive tail, highlighted by ][]{2013AJ....146...86T} in the distribution of radial peculiar velocities and Malmquist Biases give rise to a major spurious infall onto the Local Volume. This paper presents a detailed analysis of these biases and, in addition, proposes a method to retroactively derive overall better distance estimates by limiting the asymmetrical bias in radial peculiar velocity catalogs. This method results in the suppression of the large infall and thus contributes to minimize the biases responsible for this spurious general motion.

In Section 2, the biases are explained in detail. In Section 3, the {\it cosmicflows-2} features - in terms of size, data distribution and error distribution - are analyzed in order to mimic them with mock catalogs that reproduce as realistically as possible the biases. In Section 4, mock catalogs similar to the observational dataset are built out of reference cosmological simulations which resemble our Local Universe to limit the cosmic variance. A correction method is derived and tested on these mock catalogs. Three types of mocks are built to check the method outcome against the added error realization, the data selection and the constrained simulation. Namely, first type mocks differ only by the randomly added error, while mocks of the second type are constituted of a different datapoint selection from the same reference simulation. As for third type mocks, they are built out of different constrained reference simulations. After correction these mocks are input into the Wiener-Filter technique to reconstruct overdensity and velocity fields of the reference constrained simulations. Resulting reconstructions are compared with reconstructions obtained with their biased (mocks with errors and without correction) and original (mocks without errors) counterparts, and the effectiveness of the method is evaluated. Last but not least, the method is applied to the raw observational, {\it cosmicflows-2}, catalog.

Throughout this paper, distances are in \hMpc . All figures are presented after a Gaussian smoothing of (over)density fields at 2~\hMpc, the linear theory intrinsic floor value, as the WF - linear minimal variance estimator - reconstruction technique used in this paper is linear by definition. Conversions between velocity fields and displacement fields (\kms\ to \hMpc) are obtained using the linear Zel'dovich approximation \citep{1970A&A.....5...84Z} and the definition of the peculiar velocity (see the Appendix for a detailed equation). As constrained simulations were obtained within the seven-year Wilkinson Microwave Anisotropy Probe \citep{2011ApJS..192...18K} framework, reconstruction tests are conducted within the same framework.

\section{Deeper Catalogs: Stronger Biases' Effects}

Distance estimates are subject to severe systematic biases which affect the derived radial peculiar velocities. Whilst generally all gathered under the term "Malmquist Bias", three types of Malmquist Bias can in fact be distinguished. In addition to these biases, there is a lognormal error distribution which requires some attention. It is necessary to add that the biases are mentioned and explained considering the observational radial peculiar velocities catalog, {\it cosmicflows-2}, as presented in \citet{2013AJ....146...86T}. That is to say that different techniques over the years, such as those detailed in \citet{1995PhR...261..271S} relying on underlying models to correct distances as a whole before deriving radial peculiar velocities, constitute other options. Still radial peculiar velocities in the {\it cosmicflows-2} catalogs were derived without considering all the biases as a choice to privilege individual distance measurements over an ensemble of distance measurements \citep{2013AJ....146...86T}. In the four  following points, the different biases are disentangled to identify those left in the {\it cosmicflows-2} observational catalog which lead to a spurious infall onto the reconstructed Local Volume. 

\begin{itemize}
\item The most often mentioned Malmquist Bias, referred to as Problem I, Selection Effect/Bias, "r against V", Distance-dependent, Frequentist, Calibration problem, M-bias of the second kind \citep[Kaptney, 1914; Malmquist, 1922;][]{1992ApJ...395...75H,1994ApJ...430....1S,1997ARA&A..35..101T,1993A&A...280..443T,1990A&A...234....1T,1994ApJ...435..515H,1994ApJS...92....1W,1995ApL&C..31..263T}, is analogous to a selection effect in magnitude resulting in underestimated distances. A magnitude limit in a selected sample results in a mean apparent magnitude smaller than it should be. As the limit decreases, in unit of magnitudes with the distance, the bias increases. Namely, dwarfs and other dimmer galaxies are more and more under-represented in the sample with respect to brighter galaxies as the observer looks farther and farther. Consequently, their contribution to the mean apparent magnitude is increasingly reduced. Thus, the mean increases slower than it would have, with the distance or redshift, would all the galaxies have really been included. Such a bias is usually already taken care of when calibrating distance indicators. This is the case with the observational {\it cosmicflows-2} catalog \citep[e.g.][]{2012ApJ...749...78T,2013ApJ...765...94S,2014MNRAS.444..527S}. 
\item The second bias is called Homogeneous Malmquist Bias and gathers terminology such as Problem II, General Malmquist Bias, Geometry Bias, "V against r", Classical, Bayesian, Inferred-distance problem, M-bias of the first kind \citep[Kaptney, 1914; Malmquist, 1920;][]{1988ApJ...326...19L,1992ApJ...395...75H,1997ARA&A..35..101T,1994ApJ...430....1S,1993A&A...280..443T,1990A&A...234....1T,1994ApJ...435..515H,1995ApL&C..31..263T,1995PhR...261..271S}. Because our observations are restricted to a sphere centered on us, from the homogeneity of the Universe, the number of observable galaxies increases with the distance. Then considering a true distance $\langle r\rangle$, there is a higher probability to observe a galaxy at $\langle r\rangle+ dr$ than at $\langle r\rangle - dr$. Consequently, the probability for a galaxy which distance estimate is $r$ to have been located closer is higher than the opposite. Namely, distances are more likely to be underestimated. In practice, it is even more complicated than that because of the presence of small scale structures. The bias is additionally a function of the direction of observations leading to the third Malmquist Bias,
\item The last of the Malmquist Bias is referred to as the Inhomogeneous Malmquist Bias \citep[e.g.][]{1994ARA&A..32..371D,1994MNRAS.266..468H,1992ApJ...391..494L}. It is the result of the small scale structures, namely of the fluctuations of galaxy numbers. To illustrate this bias, let us consider a group of galaxies at $\langle r \rangle$ with null radial peculiar velocities for simplicity. Because of observational uncertainties, these galaxies are randomly scattered to the foreground and background of $\langle r \rangle$. For all galaxies at the same redshift, their estimated radial peculiar velocities obtained on either side of $\langle r \rangle$ result in an inaccurate infall towards $\langle r \rangle$ giving birth to spurious structures and flows. In other words, galaxies are more likely to be scattered from high density regions towards low density regions than the opposite. 
\item In addition to the Inhomogeneous Malmquist Bias, the fractional errors on distances have a lognormal distribution. The distribution of fractional errors is not symmetric when considering two galaxies one located closer to the observer than it should be and the other one positioned farther away. This asymmetry introduces a strong bias in the final recovered velocity field. In other words, distances, $d$, of galaxies are derived from moduli ($\mu$, apparent magnitude m minus absolute magnitude M) themselves obtained with observational data. Thus, assuming that errors in magnitudes are distributed evenly for moduli, since a logarithmic function is used to derive distances in megaparsecs from moduli $\mu$ in magnitudes, errors in megaparsecs are not distributed symmetrically around distance values, $d$ as shown by equations \ref{eq:modd}.
\begin{equation}
 \begin{aligned}
\mu = m -M \qquad\qquad\qquad\qquad\qquad\qquad\qquad\quad\\
\, \Delta \mu= \sqrt{\Delta m^2+\Delta M^2} \approx \Delta M \quad  \mathrm{symmetric\, around \, \mu} \;\, \\
d = 10^{(\mu-25)/5} \qquad \qquad \qquad\qquad\qquad\qquad\qquad\,\\
 \Delta d= (\Delta \mu \times \mathrm{ln}(10) \times d )/ 5 \quad\, \,\mathrm{asymmetric\, around\,} d  \,  \\
   \end{aligned}
\label{eq:modd}
\end{equation}

Since radial peculiar velocities, $v_{pec}$ and their uncertainties are derived from distances and their errors by the classical equations \ref{eq:vpec}, the asymmetric distribution propagates onto peculiar velocities. A posteriori, looking for a cause of the  observed negative tail in the distribution of radial peculiar velocities, the assumption of a Gaussian distribution of errors on distance moduli is validated.
\begin{equation}
   \begin{aligned}
v_{pec}= v_{mod}- H_{0} \times d\\
\Delta v_{pec}= H_0 \times \Delta d \quad\;\;\;\;
    \end{aligned}
\label{eq:vpec}
\end{equation}
where v$_{mod}$ is the velocity with respect to the Cosmic Microwave Background corrected for cosmological effects \citep{2013AJ....146...86T}.

This asymmetrical error distribution results in a bias in peculiar velocities. Large negative peculiar velocities have an abnormal large error which cannot be properly propagated in the WF because the sign of the error on a particular distance modulus is unknown.

As an example, let us consider a galaxy at 100 Mpc, with v$_{pec}$=0~\kms, H$_0$=75 km Mpc$^{-1}$ s$^{-1}$. Assuming that the distance modulus of this galaxy is obtained with the Tully-Fisher relation (the main provider of distances in {\it cosmicflows-2}), the 1-$\sigma$ uncertainty on the measurement is about 0.4 mag \citep{2012ApJ...749...78T,2013ApJ...765...94S,2014MNRAS.444..527S}. A 1-$\sigma$ error on the distance modulus gives either 34.2 or 35.8 mag instead of $\mu=35$ mag. The derived distance is either 69 or 145 Mpc,  giving a radial peculiar velocity of 2311 or -3338\kms.  As a result, a larger error is made by allocating a radial peculiar velocity of -3338\kms\ to this galaxy than when assuming 2311\kms. However, the sign of the error on the distance modulus is unknown, hence, although -3338~\kms\ is assigned to this galaxy, proportionally the same uncertainty is attributed to that peculiar velocity as it would have been if the value 2311\kms\ was given to that same galaxy. Since, the WF technique uses errors on peculiar velocities as an indication of the signal strength in the auto-correlation matrix, these velocities would be proportionally identically weighted despite the absolute error of the negative value being greater than the positive one. This asymmetrical bias contributes, together with the uncorrected for Malmquist biases, to the spurious overall large infall seen in the reconstructed velocity field. 
\end{itemize}

Reducing such biases is a complicated task. Grouping galaxies in groups and clusters might reduce these biases as recommended by e.g. \citet{1990ApJ...354...18B} and \citet{2011ApJ...736...93N}: because galaxies in the same cluster or group approximately share the same distance, averaging over galaxy distance estimates (to keep only distance estimates of groups and clusters) reduces errors on distances, thus on radial peculiar velocities, of clusters and groups by the square root of the number of estimates. Retracing the footsteps of methods proposed for instance by \citet{1999ApJ...522....1D} for the POTENT technique or by \citet{1994MNRAS.266..468H}, in addition to grouping, distances can be corrected or more appropriately, their uncertainties can be minimized. A method to reduce uncertainties on distances is presented in the following sections. Unlike the above mentioned methods, which assumed an underlying distance distribution as a prior to correct distances \citep[see also][]{1988ApJ...326...19L}, this paper proposes that a velocity distribution serves as a basis for the correction. Both the lack of knowledge regarding the true underlying distance distribution and the multiple distance-indicator nature of {\it cosmicflows-2} lead us to this choice. The method is thus based on correcting peculiar velocities first to decrease the flatness and skewness in the radial peculiar velocity distribution. The resulting distances derived retroactively from these corrected radial peculiar velocities are by extension better distance estimates. The set of corrected radial peculiar velocities and corrected distances results in reconstructions where the spurious infall due to biases is drastically decreased.

\vspace{-0.5cm}

\section{Analysis of {\it cosmicflows-2} catalog}

{\it Cosmicflows-2} is the second generation catalog of galaxy distances built by the Cosmicflows collaboration. Published in \citet{2013AJ....146...86T}, it contains more than 8,000 accurate galaxy peculiar velocities. Distance measurements come mostly from the Tully-Fisher relation \citep{1977A&A....54..661T} and the Fundamental Plane methods \citep{2001MNRAS.321..277C}. Cepheids \citep{2001ApJ...553...47F}, Tip of the Red Giant Branch \citep{1993ApJ...417..553L}, Surface Brightness Fluctuation \citep{2001ApJ...546..681T}, supernovae of type Ia \citep{2007ApJ...659..122J} and other miscellaneous methods also contribute to this large dataset though to a minor extent ($\sim 12\%$). The Cosmicflows team put lots of effort into gathering these data and ensuring that they are on a common scale, proceeding from closer to farther scales to check the consistency of the data and adjusting zero points when necessary. They are able to do so thanks to the numerous measurement overlaps between the different techniques. Finally, they polish their scale using that obtained with the Spitzer Telescope \citep{2012ApJ...758L..12S} which gives the great advantage, among others, of a 1\% all sky consistency erasing any concern of discrepancies between the northern and the southern observations. From these considerations, {\it cosmicflows-2} is assumed hereafter to consist of distance measurements on a common scale.

\begin{figure}
\centering
\includegraphics[scale=0.45]{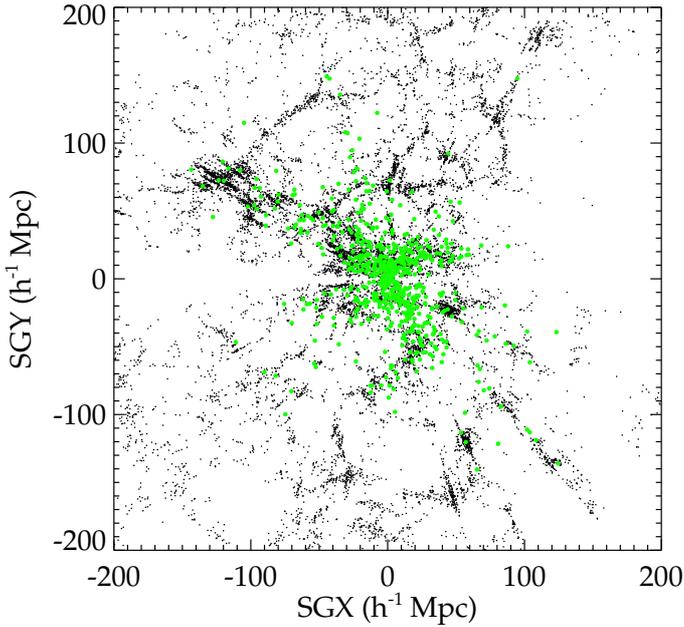}
\caption{Distribution of groups and single galaxies, in a $\pm$ 5 \hMpc\ thick slice, in {\it cosmicflows-2}, observational catalog of the Cosmicflows project (green dots) superimposed to the 2MASS redshift catalog (tiny black dots, same slice). This view (supergalactic XY plane) shows clearly the Zone Of Avoidance.}
\label{distXY}
\end{figure}

As noted in the previous section, grouping may reduced biases. Therefore, in this paper, the last grouped version of {\it cosmicflows-2} (Tully, private communication) is used as a reference to reproduce as realistic as possible biases effects in a radial peculiar velocity catalog of our neighborhood. 552 groups and 4303 single galaxies can be identified in the dataset reducing the number of radial peculiar velocities to 4855. This number is still large enough for the current scientific purpose as demonstrated by \citet{2013MNRAS.430..912D,2013MNRAS.430..902D,2013MNRAS.430..888D}. Figure \ref{distXY} shows the spatial distribution of these groups and single galaxies in a $\pm$ 5 \hMpc\ thick slice of the supergalactic XY plane on top of galaxies from the 2MASS redshift catalog within the same slice \citep[][]{2012ApJS..199...26H}. This distribution is inhomogeneous at least on two important respects: 1) the density of measurements decreases with the distance to the observer (for instance, 98 \% of the data are within a 160 \hMpc\ radius sphere centered on us but half of them are within a much smaller volume, a 60 \hMpc\ radius sphere centered on us), 2) a large zone devoid of data due to galactic extinction is visible. In order to mimic as much as possible {\it cosmicflows-2}, and as a result biases, in the related validation tests, these two features must be well reproduced by the mock catalogs. A feature which should also be mimicked comes from the error distribution. This distribution is strongly bimodal with a peak around 8-10\% fractional uncertainty on distances and another one at approximately 18 \% \citep{2013AJ....146...86T}. This bimodality is due to the different distance indicators and to the grouping: measurements obtained with supernovae, tip of the red giant branch and cepheids are given smaller uncertainties than measures obtained with Tully-Fisher or Fundamental Plane relations. As for grouping, averaging over a high number of measurements reduced the uncertainty on the measure of the group with respect to the estimate for a single galaxy.

\section{Deriving a correction method, tests on mock catalogs}
\label{mock}

Instead of correcting distances, as widely proposed in previous methods \citep[e.g.][]{1999ApJ...522....1D,1994MNRAS.266..468H}, this paper proposes first to adjust peculiar velocities and then go back to correct the distances. Consequently, the process is based on the distribution of radial peculiar velocities rather than on the radial distribution of galaxies \citep[e.g.][]{1992ApJ...391..494L,1994MNRAS.266..468H}. \citet{2001MNRAS.322..901S} proved that the distribution of radial peculiar velocities considering groups and clusters (namely removing virial motions) should be a Gaussian. Consequently, unless the Milky Way is at a peculiar position in the Universe, which is highly improbable, the distribution of radial peculiar velocities obtained from our position should be close to a Gaussian too. As a matter of fact, Gaussianity is found in mock peculiar velocity catalogs drawn from N-body simulations. Namely, dark matter halos (accessible with dark matter only simulations, themselves well described by the linear theory) are equivalent to groups, clusters or isolated galaxies. To derive a method to minimize biases in observational datasets, this Gaussianity will be the initial assumption. \citet{2008PhDT........32B} noted, when studying radial peculiar velocities in simulations, that although the overall distribution of radial peculiar velocities was confirmed as Gaussian, we still can be affected by:
\begin{itemize}
\item the cosmic variance due to the particularity of our neighborhood,
\item Poisson noise due the restricted size of the sample of peculiar velocities.
\end{itemize}
Still, a major advantage of this paper study comes from the fact that:
\begin{itemize}
\item constrained simulations of the Local Universe are used to produce a set of mocks on which the method is tested,
\item mock catalogs mimic as much as possible the characteristics of the observational catalog under study.
\end{itemize}
As a result, both cosmic variance and Poisson noise are reduced. In any case,  peculiarities of the Local Universe make purely mathematical statistical methods not optimal.

\subsection{Mock catalogs of peculiar velocities}
\begin{figure}
\includegraphics[scale=0.6]{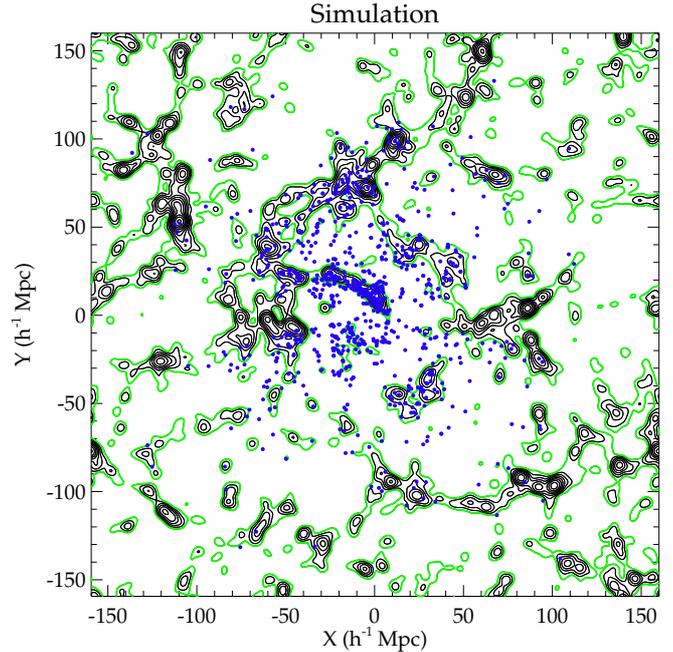}
\caption{Example of a distribution of selected halos (blue dots) in the XY plane ($\pm$ 5 \hMpc\ slice in Z) of a cosmological simulation which resembles the Local Universe.  Halos are selected to build a mock similar to the second catalog of the Cosmicflows project. Black and green contours show the density field and the mean density of the cosmological simulation.}
\label{simu}
\end{figure}

\begin{figure}
\centering
\includegraphics[scale=0.6]{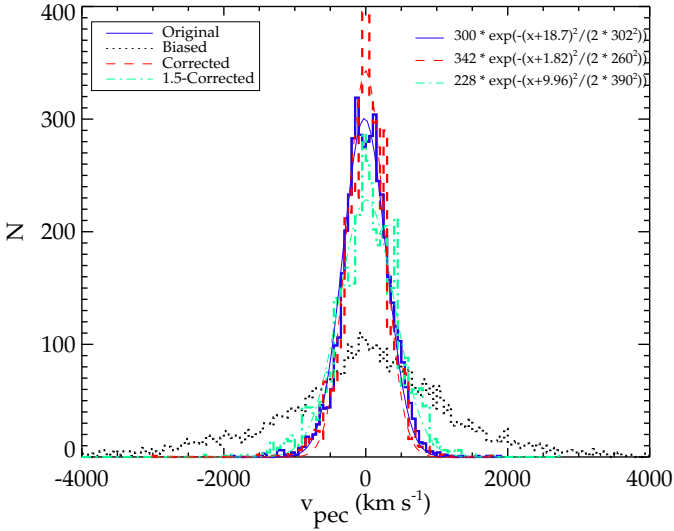}
\caption{Distribution of radial peculiar velocities in the original (without error) mock catalog (blue solid thick histogram) and in the biased mock (black dotted histogram). The original distribution of radial peculiar velocities can be modeled by a normal distribution (blue solid light curve). The distributions of corrected and 1.5 corrected radial peculiar velocities are shown by the red dashed thick and green dot-dashed thick histograms. Gaussians can also fit these distributions (red dashed light and green dot-dashed light lines). }
\label{gauss}
\end{figure}

\begin{figure}
\includegraphics[scale=0.6]{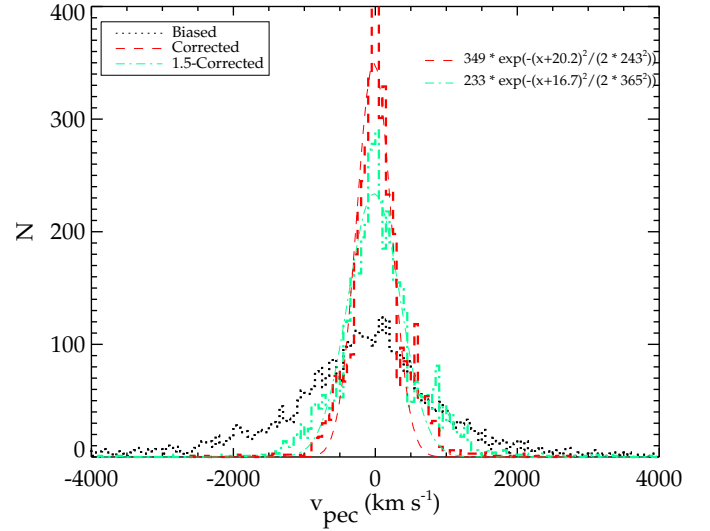}
\caption{Radial peculiar velocity distributions in CF2-Biased (black dotted line), CF2-Corrected (red dashed thick line) and CF2-1.5-Corrected (green dot-dashed thick line) catalogs. A larger tail is visible in the negative side of this diagram due to the error bias in the velocity distribution of CF2-biased. After correction, the distributions of radial peculiar velocities in {\it cosmicflows-2} corrected datasets can be fitted by Gaussians (red dashed and green dot-dashed light lines) with parameters close to the theoretical ones (obtained with mocks). }
\label{symasym}
\end{figure}

To limit the impact of the cosmic variance in the following tests, simulations resembling the Local Universe down to the linear threshold (2 \hMpc) are chosen. Our simulations have 512$^3$ particles and are 320 \hMpc\ wide. They were computed within the framework of the CLUES\footnote{http://www.clues-project.org/} project \citep[Constrained Local Universe Simulations,][]{2010arXiv1005.2687G} using the method described in \citet{2014MNRAS.437.3586S}. Unlike typical simulations, constrained simulations stem from a set of constraints that can be either redshift surveys or radial peculiar velocities. Simulations used here obey a set of radial peculiar velocities. Most importantly, a look alike for all the major structures and voids of the Local Universe can be found in these simulations ensuring that the Large Scale Environment is similar to our neighborhood. Several realizations are tested to measure the robustness and the accuracy of the method discussed in the next subsection. Results are similar, if not identical, for every mock tested. Consequently, while the figures are found by averaging over the entire mock set, one mock (chosen randomly) built out of one of the aforementioned simulations is presented in this paper in terms of plots. Considering the box to be equivalent to the Local Universe, we place an observer at the center of the box, coordinates can be defined similarly to observational supergalactic coordinates. The XY plane in this set of coordinates, is shown in Figure \ref{simu}: a Shapley supercluster candidate in the top left corner, a look-alike of Coma in the top middle, and a candidate for Perseus-Pisces around [-25,25] \hMpc, Virgo's look-alike is close to the center and the Centaurus-Great Attractor region-like is on Virgo's left side. Using Amiga halo finder \citep[][]{2009ApJS..182..608K}, a list of halos is drawn from this simulation. Halos are then selected and prepared to match {\it cosmicflows-2} grouped catalog in three steps:
\begin{itemize}
\item We seek to have a similar repartition of data points (number, spatial coverage and distribution including the Zone Of Avoidance). Subsequently, every halo in a zone similar to the Zone of Avoidance is removed from the list. This zone is defined as a cone with the apex at the center of the box (where the observer is assumed to lie) and assuming the same orientation within the XYZ volume as the observational one in the supergalactic XYZ volume. Then, an histogram with bin size of 20~\hMpc\ is derived for the observational {\it cosmicflows-2} catalog providing the number of measurements in each bin or 20 \hMpc\ slice. For each 20 \hMpc\ slice, the same number of halos as found in {\it cosmicflows-2} is selected randomly in the list of halos. In case of multiple halos in the same region, the most massive halo is privileged to mimic the grouping applied to {\it cosmicflows-2}. On Figure \ref{simu}, the resulting compiled list of halos is visible as blue dots in a 10 \hMpc\ thick slice in the XY plane. As predicted by \citet{2001MNRAS.322..901S}, before inserting errors, the halo radial peculiar velocity distribution (computed with respect to the box center, where the Milky-Way like is assumed to be) of this mock catalog can be modeled by a Gaussian visible in Figure \ref{gauss} in blue. This mock is called hereafter original in the sense that it has no error on galaxy distances yet.
\item Next, we add errors to radial peculiar velocity distributions preserving the asymmetry problem. Accordingly, a Gaussian distribution of errors with 0.17 magnitude scatter (as on average by assumption in the observational catalog) is added to distance moduli (not to distances) and distorted distances and corresponding radial peculiar velocities are computed. Because distances are distorted the Malmquist biases are also reproduced. Figure \ref{gauss} displays the distribution of radial peculiar velocities for this mock with errors, hereafter biased mock, by a black dotted line. This distribution is flatter than a theoretical Gaussian with a slightly larger tail on the negative side confirmed by a negative skewness value. This distribution and that of {\it cosmicflows-2} represented also by a dotted black line but on Figure \ref{symasym} are similar to each other confirming the 0.17 mag chosen scatter. 
\item Finally, the inserted errors are the 'true errors' but are unknown to observers. Observers have at their disposal only the  1-$\sigma$ uncertainties which differ on the distance indicators and the grouping. This fractional distribution of 1-$\sigma$ uncertainties is globally highly bimodal in {\it cosmicflows-2} \citep{2013AJ....146...86T}, accordingly distances are assigned in majority a 17\% and in minority a 5\% 1-$\sigma$ fractional uncertainty. This assignment is not done entirely randomly but in accordance with the real inserted errors in the previous step, analogously to how observers attribute the fractional uncertainty in function of the distance indicator.
\end{itemize}

We repeat this three-step process several times with this and all the other constrained simulations at our disposal. F-tests show that resulting radial peculiar velocity distributions are always similar at the 98 $\pm$ 2 \% confidence level. 
 
\subsection{Method: minimization of the Biases using Mocks}

We intend to present a method to correct overall distance estimates by correcting first radial peculiar velocities. Figure \ref{symasym} reveals a skewness towards negative peculiar velocities in {\it cosmicflows-2} confirming the asymmetry phenomenon. However, \citet{2001MNRAS.322..901S} proved that the radial peculiar velocity distribution should be a Gaussian. Consequently, to derive a correction for observational datasets, this Gaussianity is the initial assumption in the method proposed in this paper.

After application of the correction-process, catalogs should have a Gaussian radial peculiar velocity distribution with variance and location peak similar to that of the original ones. In mock cases, variance and peak are determined by mocks with original positions, hence velocities. Regardless, in the different original mocks, standard deviations of Gaussians fitted to radial peculiar velocity distributions are always around 300 $\pm$ 50 \kms. Actually, F-test results, given on average in Table \ref{Tbl:1}, column 6, reveal that a Gaussian of variance 300 \kms\ can be accepted on average as a model for original radial peculiar velocity distributions at the 98 $\pm$ 2 \% confidence level. As a result, a biased peculiar velocity is modified according to its probability of belonging to the theoretical Gaussian (with a typical standard deviation of 300~\kms) and according to its uncertainty. Two cases can be distinguished, either the radial peculiar velocity is positive or it is negative. Then corrected radial peculiar velocities ($v_{pec\ c}$) are derived with equations \ref{eq:1} and \ref{eq:2} devised in this work. 
\begin{equation}
\mathrm{if}\, v_{pec}>0, \, v_{pec\, c}=w\ [ p  (v_{pec} - \Delta) + (1-p)  (v_{pec} +  \Delta)] + (1-w)\ v_{pec}
\label{eq:1}
\end{equation}
\begin{equation}
\mathrm{if}\, v_{pec}<0, v_{pec\, c}=w\ [ p  (v_{pec} +  \Delta) + (1-p)  (v_{pec} -  \Delta)] + (1-w)\ v_{pec}
\label{eq:2}
\end{equation}
where $\Delta$ is the radial peculiar velocity uncertainty ($\Delta v_{pec}$). $p$ and $w$ are two weight factors between two quantities. More precisely, $p$ is the probability that a radial peculiar velocity does not belong to the theoretical Gaussian (thus it needs to be corrected and it should either be reduced if it is highly positive or increased it is highly negative). $p$ acts as the weight factor between $v_{pec} - \Delta$ and $v_{pec} + \Delta$ to establish the quantity by which a velocity is decreased/increased. This gives an intermediate velocity value but no information regarding the uncertainty on the velocity estimate has yet been considered. To use this additional information, $w$ is the weighted uncertainty in opposition with $\Delta$ ('unweighted' uncertainty). $w$ is the weight between the intermediate value and the measured value.
The weight factors are derived with the following equations:
\begin{equation}
p=1- \int_{-\infty}^{\, -|v_{pec}|} exp(-\frac{v^2}{\sigma^2})dv \qquad w=\frac{\Delta}{H_0 d} \times (\mathrm{max}(\frac{\Delta}{H_0 d})+0.08)^{-1}
\end{equation} 
The 0.08 term in $w$ is required to ensure that we keep a minimum of trust even towards radial peculiar velocities with the maximum (max) fractional uncertainty. Several values between 0 and 1 were tested, 0.08 being found to be the best parameter to retrieve a distribution close to the theoretical Gaussian (at the 80 $\pm$ 3 \% level) at the end of the correction process. In other words, to correct velocities, their uncertainty (weighted, $w$ or not, $\Delta$) and their `position' on the theoretical Gaussian ($p$) are the two parameters used.

As $\Delta$ corresponds to the 1-$\sigma$ uncertainty, an iterative process was necessary. Namely, after the first iteration, peculiar velocities which are still in the 1-$\sigma$ and beyond range of the theoretical Gaussian distribution are input again in the equations. After, this second iteration, velocities in the 2-$\sigma$ and beyond range are corrected again. Four iterations are sufficient to retrieve the Gaussian distribution without any long (negative and positive) tails and skewness. For consistency, corresponding distances need to be computed to correct for the Malmquist biases. They are derived using corrected radial peculiar velocities and the classical formula \ref{eq:vpec} in reverse order. \\

\begin{figure}
\includegraphics[scale=0.55]{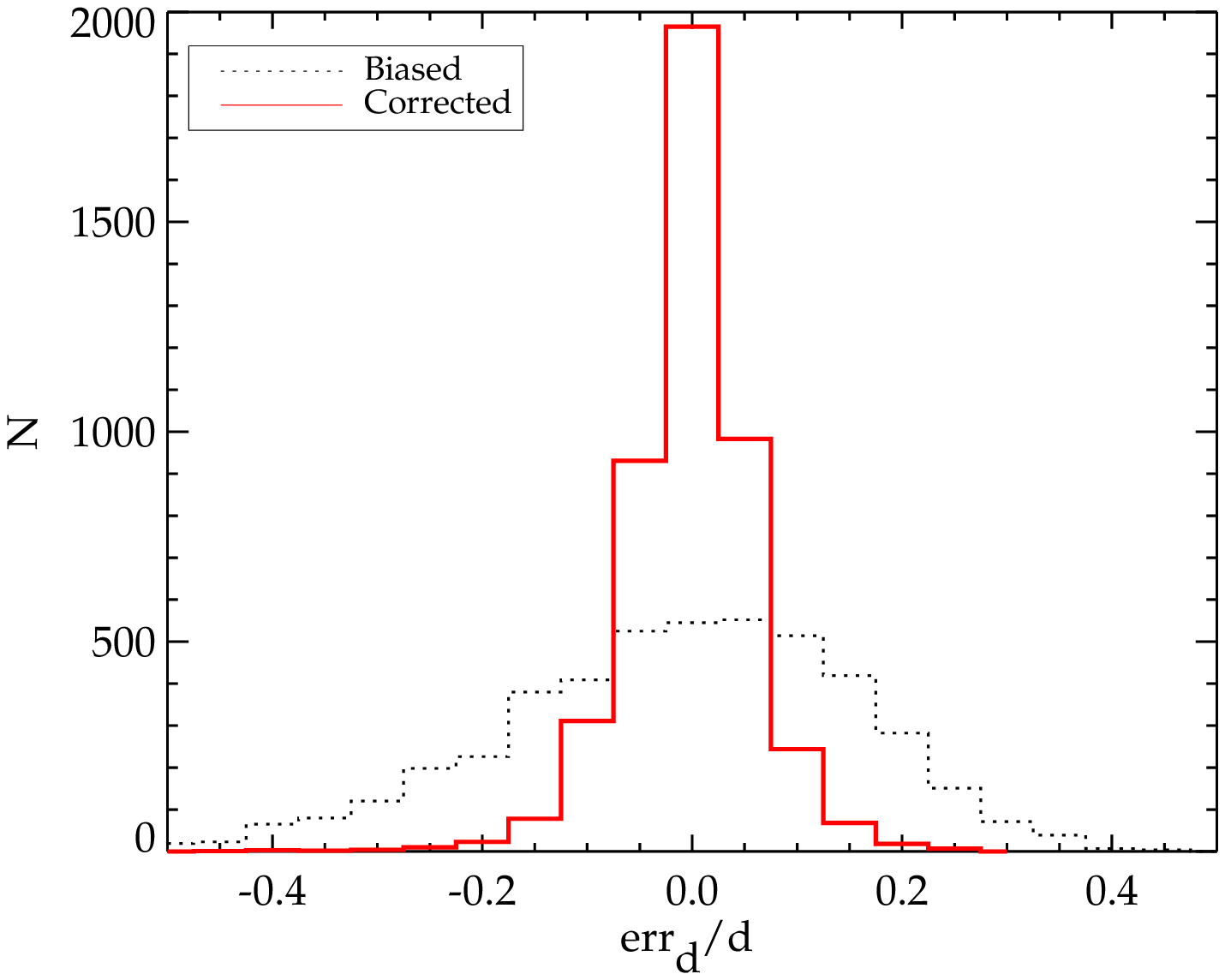}\\
\hspace{-0.15cm}\includegraphics[scale=0.55]{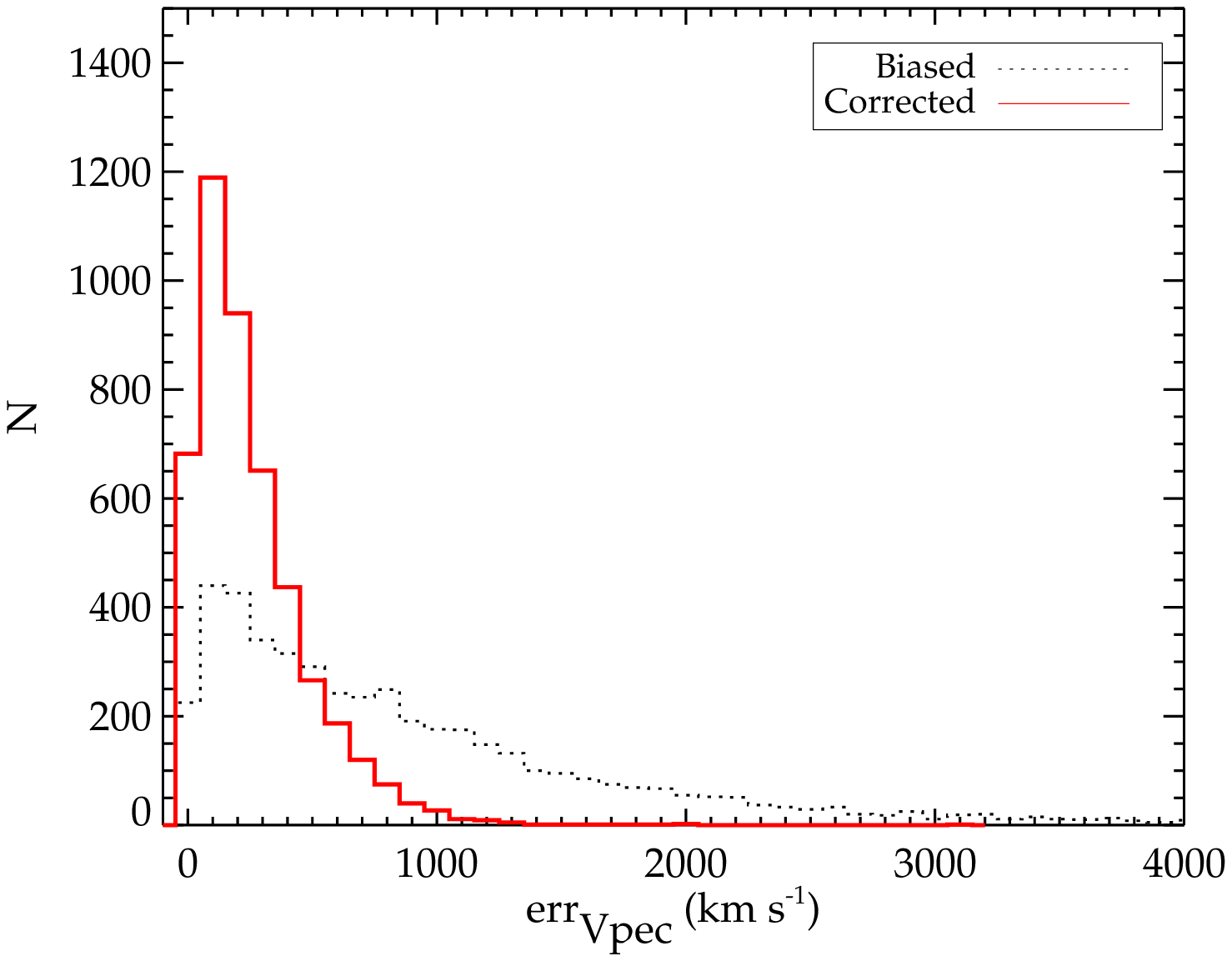}
\caption{Histograms of fractional errors on distances (top) and absolute errors on radial peculiar velocities (bottom) in the biased mock (dotted black lines) and in the corrected mock (solid red lines). Absolute fractional errors on corrected distances have a median at 0.03 and a mean at 0.04.}
\label{beforeafter}
\end{figure}

After correction, the distributions of radial peculiar velocities, fractional errors on distances, and absolute errors on radial peculiar velocities are all shown on Figures \ref{gauss} and \ref{beforeafter} in red for the mock selected for display. 
Figure \ref{beforeafter} also presents, before correction, the histograms of fractional errors on distances (top) and of absolute errors on radial peculiar velocities (bottom) by black dotted lines. Absolute fractional errors on corrected distances have a median at 0.03 and a mean at 0.04. Fractional errors on distances are distributed on an approximate Gaussian and the distribution of uncertainties on radial peculiar velocities is less flat and contained in a smaller interval of values. These distributions, mean and median values are typical for all the mocks built out of the different constrained simulations. Distribution of radial peculiar velocities can now be approximated by a Gaussian with variance 300~\kms\ at the average confidence level of 80 $\pm$ 3 \% according to F-tests. On the other hand such a Gaussian can be rejected as a model at the average 98 $\pm$ 2 \% confidence level for radial peculiar velocity distributions of biased mocks. The average F-test values are gathered in Table \ref{Tbl:1}, columns 7 and 8. 

\begin{figure}
\hspace{0.25cm}\includegraphics[scale=0.5]{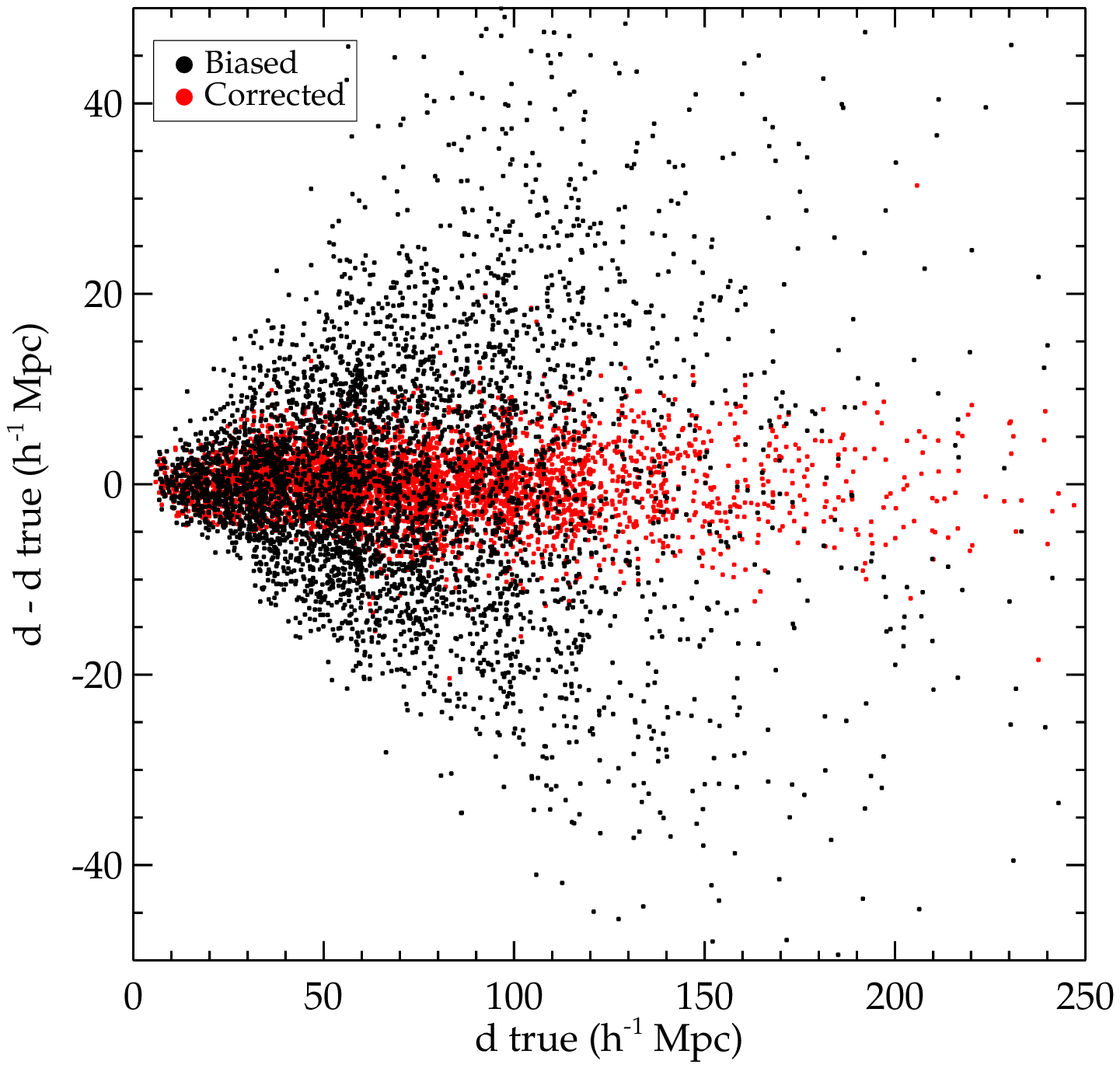}\\
\includegraphics[scale=0.5]{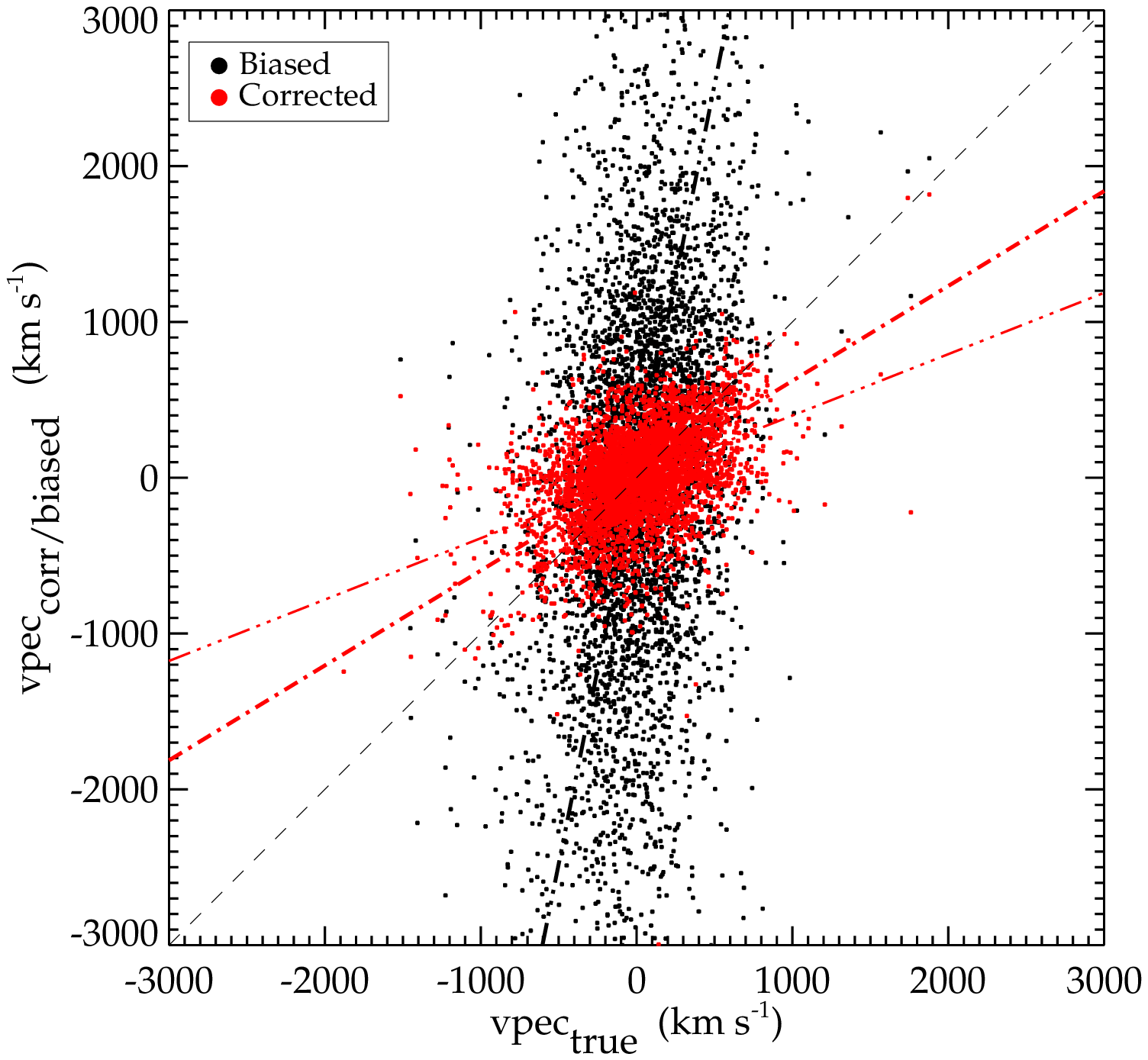}
\caption{Top: Residual between biased (black), corrected (red) distances and true distances versus true distances. The standard deviation is about 15 \hMpc\ before correction while it is only about 4 \hMpc\ after correction. Bottom: Biased (black) and corrected (red) velocities versus true velocities. The thin black dashed line shows the ideal correlation. The thick black dash-dotted line is the correlation before correction while red lines stand for correlations after correction. The thinner, triple dots-dashed, red line shows the correlation without assuming any special errors on velocities while the thicker dotted-dashed red line shows the correlation assuming a 0.04 fractional error on corrected distances.}
\label{diffcorr}
\end{figure}

Figure \ref{diffcorr} is a proof of concept that corrected distances and corrected radial peculiar velocities are overall closer to their true values than biased ones. In the top panel of this figure, residuals between erroneous and true distances are drastically reduced when comparing corrected and true distances instead of biased and true distances. The standard deviation decreases from about 15 to 4 \hMpc. As for velocities, the correlation between corrected and true velocities is closer to the ideal one than without correction.
Moreover, no additional bias is created since, corrected distances are not always higher/smaller than original ones. Namely, distances are not all under/over-estimated. The reconstruction technique is put into practice in the next subsection on the minimized biased mocks. Fractional errors of 4\% on distances are assumed, in agreement with the upper limit of fractional error medians found in the various corrected mocks. As shown in Figure \ref{diffcorr}, this assumption improves the correlation between corrected and true velocities.

\subsection{Reconstruction using Mocks}

For purposes of completeness, to check the robustness of the correction method and the independence of its outcome on the error realization, the halo selection and the constrained simulation, the following selections are made.
\begin{itemize}
\item Ten mock catalogs built out of the same distribution of halos are randomly selected. They differ only by the realization of errors. They are denoted errorMock. Namely, error distributions are still Gaussians with 0.17 magnitude scatter but built out of different seeds such that different errors are attributed to the same halo in two different mocks of this series ;\\
\item Ten mock catalogs built out of the same reference simulation are randomly selected. They, hereafter haloMock, differ by the selection of halos. Namely, halo distributions are still mimicking {\it cosmicflows-2} but different halos are randomly picked to build the mocks ;\\
\item Five mock catalogs out of five different realizations/reference simulations are randomly selected. They differ by the random component used to build the constrained initial conditions. These five different realizations, hereafter referred to as simuMock, were chosen to resemble the Local Universe but also to present a variety of velocity field monopole term (term responsible for an infall if highly negative or an outflow if highly positive). The importance of this term will be developed later in this paper.
\end{itemize}
 
\begin{figure*}
\flushleft
\includegraphics[scale=1.2]{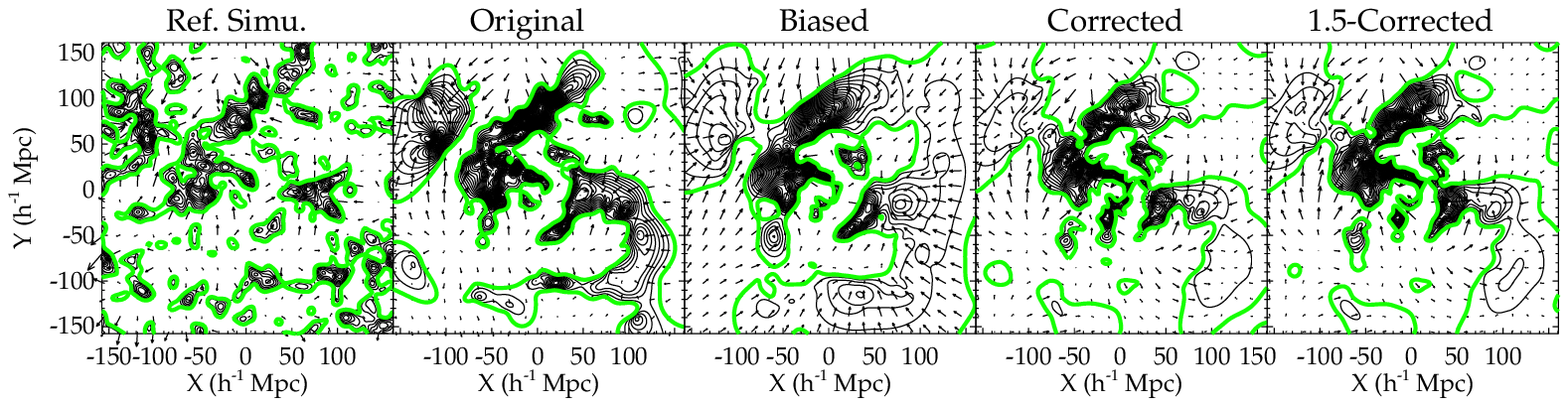}
\includegraphics[scale=1.2]{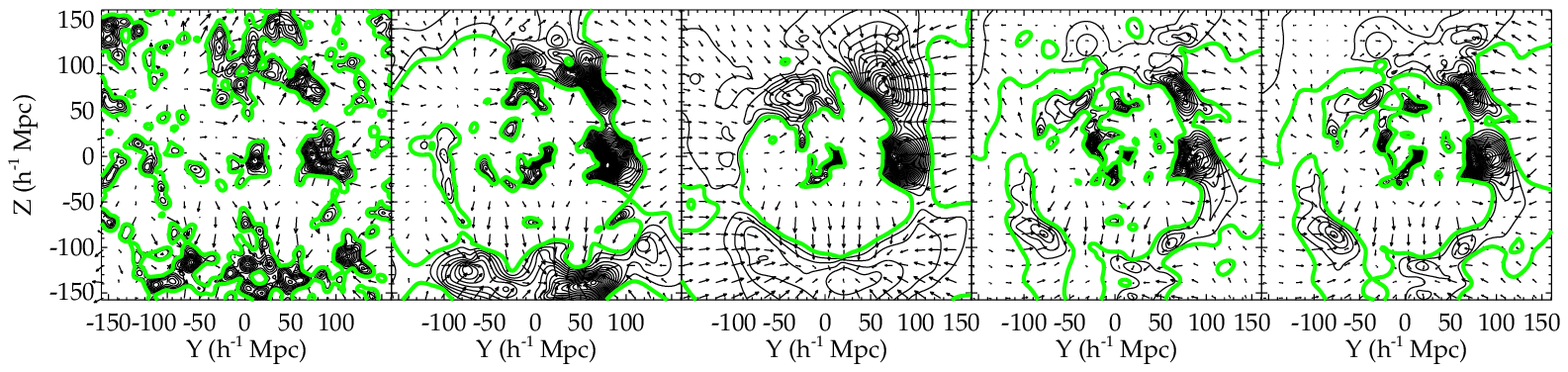}
\includegraphics[scale=1.2]{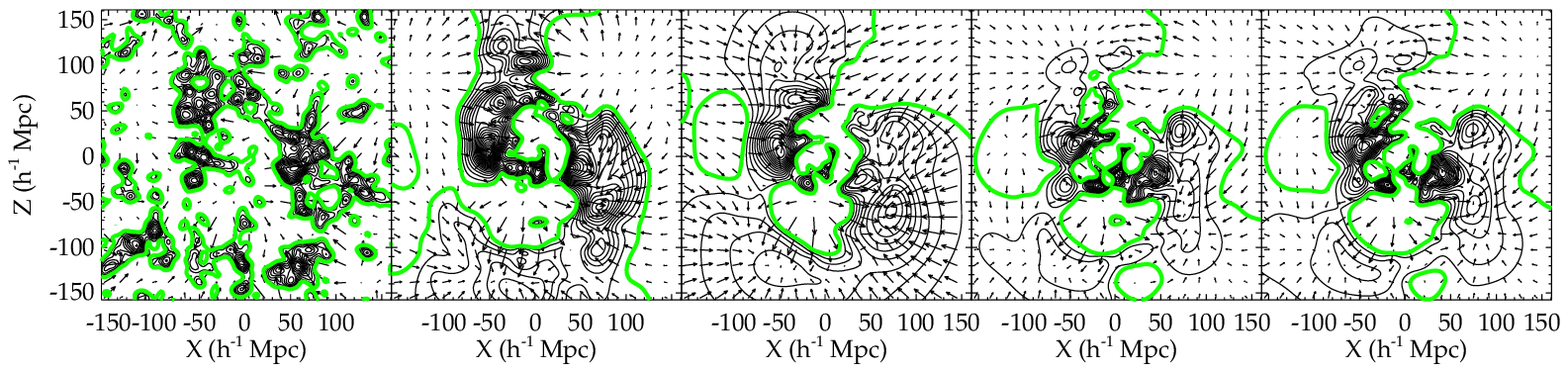}
\caption{XY (top), YZ (middle) and XZ (bottom) slices of the (over)density (black contours) and velocity (black arrows) fields from the reference simulation (left) and reconstructed with the Wiener-Filter technique applied to the original, ie. without error (middle left), biased (center), corrected (middle right) and 1.5-Corrected (right) mocks. Green contours show the mean density. The net spurious infall onto the volume is clearly visible in the velocity field reconstructed from the biased mock and structures are round. On the opposite, structures are more sharply defined in the reconstructions obtained with the corrected and 1.5-Corrected mock and the strong infall has disappeared (for instance, the expulsion from the void in the top right corner of the XZ plane is recovered) in agreement with the best reconstruction possible, of the reference simulation (first column), with a mock similar to {\it cosmicflows-2} but without error (second column). No major difference can be seen between the results obtained with the corrected and 1.5-Corrected in agreement with the fact that they differ only in terms of the divergent part (velocities due solely to densities in the box) of their velocity fields.}
\label{WFall}
\end{figure*}

Next, the WF technique is applied to each one of the mocks, as described above, plus their corrected versions and the original versions. In other words, the WF is applied to: 1) the biased catalog (either an errorMock, a haloMock or a simuMock),  2) the corresponding corrected catalog (the biased mock corrected with the method) and, as a control, 3) the catalog with original positions and radial peculiar velocities, to remove any bias due to an increasing smoothing with distance. This sample of 66 mocks (25 biased mocks, 25 corrected mocks and 16 original mocks, the ten mock catalogs built out of the same distribution of halos have the same original mock) are all used to reconstruct the overdensity and velocity fields of the reference simulations which resemble the Local Universe. Reconstructions obtained with the original mocks represent the best overdensity and velocity fields we can expect from the WF method for a given reference simulation and halo selection. As the goal of the paper is not to test the accuracy of the WF technique, already widely tested \citep[e.g.][]{1994ASPC...67..185H,1995ApJ...449..446Z,1999ApJ...520..413Z,1999elss.conf..148H,2000ASPC..218..173Z,2012ApJ...744...43C,2014MNRAS.437.3586S}, it is in a sense easier to compare reconstructions obtained with biased and corrected mocks to those obtained with original mocks to determine to which extent biased mocks are properly corrected. For completeness however, the reference simulation, selected for plots in this paper, is shown with the three reconstructions on Figure \ref{WFall}. The reference simulation, from which the mock catalog was extracted, is plotted in the first column. The three reconstructions obtained with the WF applied to the three different mocks (original, biased and corrected) are shown in the next three columns. The reconstruction obtained with the biased mock (third column) presents what is expected from biases' effects, namely a large infall in contradiction with the reference simulation (first column). In addition, structures are extended and very round. On the other hand, the strong infall has disappeared from the reconstruction resulting from the corrected mock (fourth column). Structures are more sharply defined in good agreement with the best result we can obtain using the WF technique on a catalog similar to {\it cosmicflows-2} but original (second column), namely without errors. We emphasize that the WF reconstruction obtained with this original mock is that which reconstructs the best the reference simulation for a given halo selection. Other reconstructions (from mocks with errors) cannot be any better.\\

After this qualitative analysis, a first quantitative analysis can be directed to the aspect of the infall. The concept is to compute monopole terms at 150 \hMpc\ of the WF reconstructed velocity fields, terms which are proportional to the divergence of these latter. Decreasing (in absolute value) the monopole term  \citep[trace of the symmetric component of the deformation tensor, e.g.][]{2013MNRAS.428.2489L,2014MNRAS.441.1974L} is equivalent to reducing the infall onto the Local Volume. This infall is due to biases and as such is unphysical. The limit of 150 \hMpc\ is the result of the assumption that, at such a distance, the effect of radial peculiar velocities on galaxy total velocities has to be low. The expansion is predominant at such distances. As a matter of fact, monopole terms of velocity fields obtained with the WF technique applied to corrected mocks happen to be divided approximately by a factor ten in comparison with those of velocity fields obtained with biased mocks. Table \ref{Tbl:1} gathers the average monopole terms for the three different types of mocks as well as for the entire set of mocks. On Figure \ref{WFall}, the infall observed in the displayed reconstruction obtained with the, selected for display, biased mock is well reflected by the average large negative monopole term value (column 3: -792 $\pm$ 71 \kms). On the other hand, original and corrected mocks result in reconstructions without a large infall in agreement with low monopole term values found for the corresponding reconstructed velocity fields and for that of the reference simulation (columns 2 and 4: -70 $\pm$ 47 \kms\ and -58 $\pm$ 42 \kms). The small scatters show that it is a general trend. Monopole term values are more similar between velocity fields obtained with original and corrected mocks than with biased mocks. The largest difference between monopole term values of velocity fields from reference simulations - or obtained with the WF applied to original mocks - and those obtained with corrected mocks is observed for the two simulations with less `Local Universe-like' values (quite large absolute values, -210 and 210 \kms), thus they are supposed to be less representative of the Local Universe. As for the most common difference values, they are of the same order of magnitude ($\sim$ 20-30 \kms) as the value variations found for velocity fields obtained with original mocks built from the same reference simulation but consisting of a different halo selection (haloMock category).

\begin{figure*}
\includegraphics[scale=0.39]{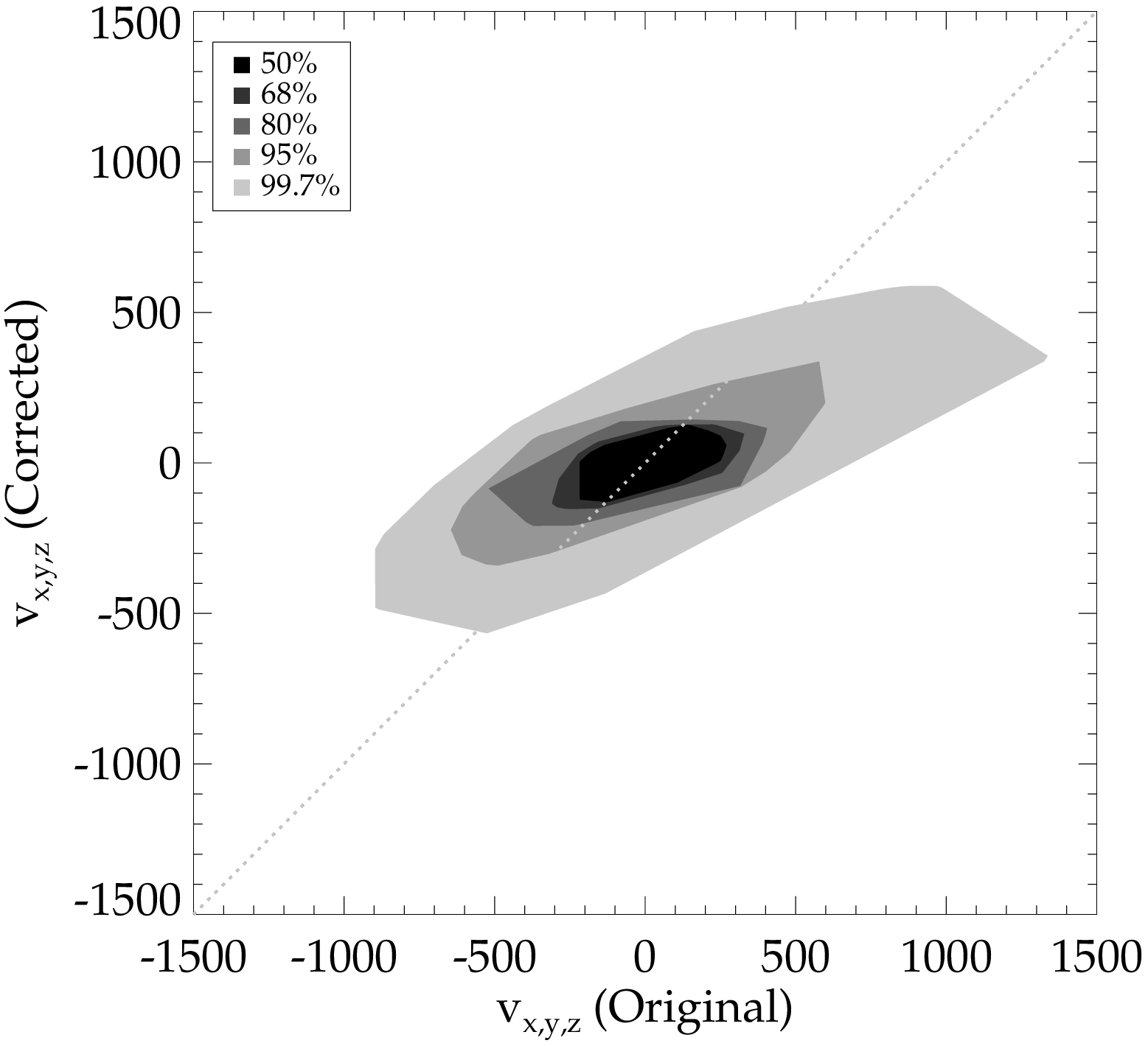}
\includegraphics[scale=0.39]{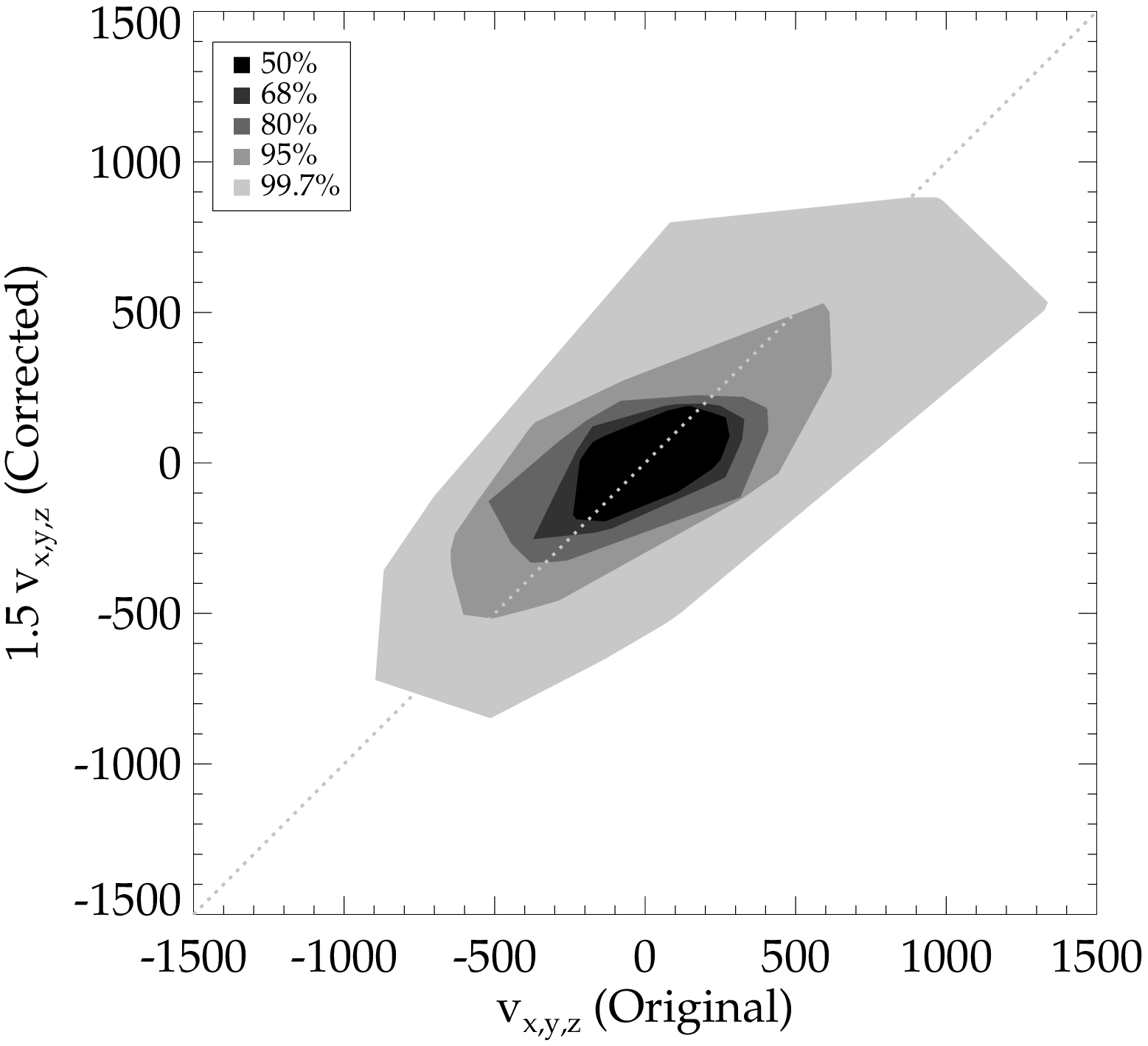}
\includegraphics[scale=0.39]{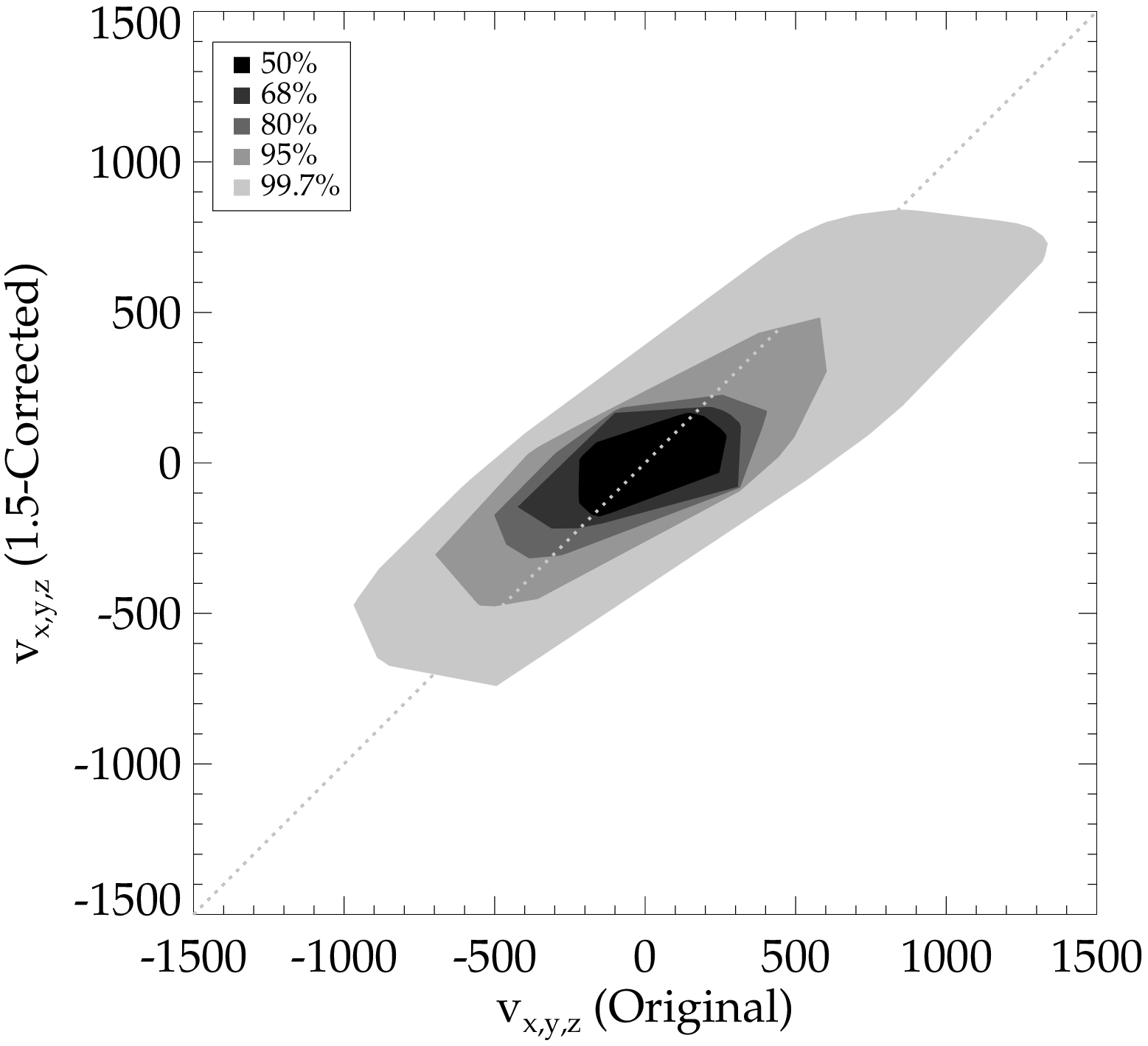}	
\vspace{-0.25cm}
\caption{Cell-to-cell comparisons within 320 \hMpc\ between the Wiener-Filter reconstructed full velocity fields obtained with original and corrected mocks (from left to right, Corrected, Corrected but reconstructed velocities multiplied by 1.5, 1.5-Corrected versus original). The gradient of grey delimits the 50, 68, 80, 95 and 99.7 \% confidence zones. The tilt with respect to the 1:1 linear relation is decreased in the cell-to-cell comparisons of the Wiener-Filter reconstructed full velocity fields obtained with the corrected mock but reconstructed velocities multiplied by 1.5 and with the 1.5-Corrected mock versus the original. The scatter is smaller for the comparison between full velocity fields obtained with the 1.5-Corrected and original mocks.}
\label{c2c1}
\end{figure*}

A second quantitative analysis can be derived from cell-to-cell comparisons between reconstructed velocity fields obtained when applying the WF technique to original and other mocks. In Table \ref{Tbl:1}, columns 11 and 12, 1-$\sigma$ scatters are smaller, by $\sim$ 10 - 20 \kms, when comparing grid cells within 320 \hMpc\ of the WF reconstructed velocity fields obtained with original and corrected mocks than those obtained with original and biased mocks. However, on Figure \ref{c2c1}, left-hand panel, because the WF smooths by definition, a tilt can be observed in the cell-to-cell comparison of reconstructed velocity fields obtained with corrected and original, chosen for display, mocks. Such a tilt can be observed in every comparison between a velocity field obtained from a corrected mock and its corresponding original mock's WF velocity field. Additional cell-to-cell comparisons between reconstructed divergent (due solely to densities in the box) velocity fields reveals that the correction combined with the WF technique affects divergent velocity fields, which are not affected as much as full (divergent plus tidal) fields by biases as shown in Table \ref{Tbl:1}, columns 11 and 12. In other words, input velocities have to be high enough in order to be sufficiently high in the final reconstruction because of the Wiener-Filter smoothing. Currently, reconstructed velocities are too low in absolute value by an average factor of 1.5. A test was conducted accordingly. WF resulting velocities are multiplied by 1.5 in all the reconstructions obtained with corrected mocks. In every case, 1-$\sigma$ scatters are decreased some more (by $\sim$ 5 \kms) in cell-to-cell comparisons between adjusted by 1.5 velocity fields obtained with corrected mocks and those obtained with original mocks. Because WF reconstructed velocity fields abide by a prior (cosmology), this test serves the sole purpose of demonstrating that the tilt with an empirical 1.5 factor is a constant. Looking back to Figure \ref{diffcorr}, the linear fit obtained assuming a 0.04 fractional error on corrected distances gave us a hint about this 1.5 factor as, for instance, a true radial peculiar velocity of 2000 \kms\ correlates via the fit to a corrected radial peculiar velocity of 1000 \kms. Then, to remedy to the problem, corrected velocities are multiplied by 1.5 \emph{before} the reconstruction and fractional errors are re-adjusted to 5\% before applying the WF on the 'new' corrected mocks, hereafter denoted 1.5-Corrected. 

The last column of Figure \ref{WFall} shows the WF reconstruction obtained out of the randomly selected for display 1.5-Corrected mock. Comparing this reconstruction with the reconstruction obtained with the corrected mock given in the fourth column of the same Figure reveals no major qualitative differences. Quantitatively, Figure \ref{c2c1} shows that no evident tilt can be observed anymore when comparing the WF full reconstructed velocity fields obtained with original and 1.5-Corrected mocks. Again, in Table \ref{Tbl:1}, 1-$\sigma$ scatters are smaller, by 20 - 30 \kms, when comparing full velocity fields reconstructed with the WF applied to original and 1.5-Corrected mocks than applied to original and biased mocks. Monopole terms (in absolute value) computed for these reconstructed velocity fields are also decreased with respect to velocity field monopole terms obtained with biased mocks (by a factor $\ge$ 5) . The average monopole terms are also gathered in Table \ref{Tbl:1}, column 5. The additional comparisons between reconstructed divergent velocity fields reveal that the correction combined with the WF technique no longer affect the divergent velocity field as scatters are either unchanged or slightly decreased. The only (although non significant) exception appears for one mock of the errorMock category but this mock is not representative of {cosmicflows-2} because of a 'relatively low' monopole term (-510 \kms) in the reconstruction obtained with the biased mock (namely a smaller bias effect). As a matter of fact applying the WF to \emph{cosmicflows-2}, hereafter CF2-biased, we found a monopole term value at 150 \hMpc\ close to -800 \kms\ for the resulting reconstructed velocity field. 

\begin{figure*}
\centering
\vspace{-0.1cm}
\hspace{-1.25cm}\includegraphics[scale=1]{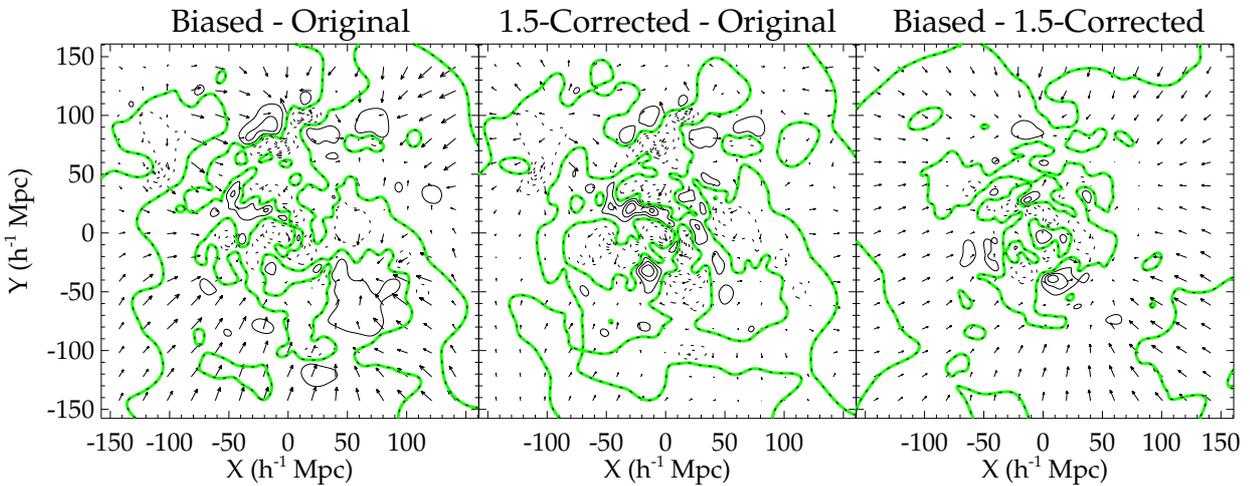}
\caption{XY slices of residuals between overdensity (black and dashed contours for positive and negative difference value) and velocity (black arrows) fields reconstructed with the Wiener-Filter technique applied to the biased (left), 1.5-Corrected (middle) and original mocks and to the biased and 1.5-Corrected mocks (right). Arrow lengths are proportional to velocity residuals. Green contours separate positive and negative density residuals. The pattern of infall present in the differential plot between reconstructed velocity fields obtained with biased and original mocks is absent in the differential plot between velocity fields reconstructed with 1.5-Corrected and original mocks. No additional patterns appear. Differences in structures are similar in both cases: neither major structures, nor major voids are created by the correction. The right panel confirms that the reconstruction obtained with the 1.5-Corrected mock is close to the best possible reconstruction (from the original mock) as the residual between velocity fields obtained with biased and 1.5-Corrected mocks resembles that observed for velocity fields reconstructed with biased and original mocks.}
\label{diffplots}
\end{figure*}

Differential plots allow another verification which consists in checking that no patterns are created by the correction in WF reconstructions. These plots represent the difference or residual between two reconstructions. Figure \ref{diffplots} shows such plots for the XY slice of the WF reconstructed overdensity and velocity fields obtained with the, selected for display, biased, corrected and original mocks. Arrow lengths are proportional to the residual of the velocity fields, dashed and solid black contours represent the residual of the overdensity fields. Green contours separate positive and negative residuals. From the two first columns, the difference between velocity field reconstructions from biased/original (left) and corrected/original (middle) mocks clearly confirms that the infall is suppressed by the correction, the residual is decreased and no pattern is created. Numerically, the root mean squares of velocity field residuals (sum in quadrature of $vx_1-vx_2$, $vy_1-vy_2$, $vz_1-vz_2$ components where 1 and 2 stand for WF reconstructed velocity fields obtained either with biased, original or 1.5-Corrected mocks) are 350 \kms\ for biased/original mocks and 250 \kms\ for 1.5-Corrected/original mocks. For comparison, the root mean square of the velocity field residual for two original mocks from the same reference simulation (haloMock category) is about 150 \kms. No created structures are evident in the density field residuals, confirming the absence of 1-$\sigma$ scatter changes in cell-to-cell density field comparisons, gathered in Table \ref{Tbl:1}. The small variation in the 1-$\sigma$ scatters obtained with cell-to-cell comparisons of divergent velocity fields (derived from densities) constitutes another proof. On that same Figure, velocity field residuals, between reconstructions obtained with biased/original (first panel) and biased/1.5-Corrected (last panel) mocks, are similar. This highlights again that the correction results in reconstructions close to the best possible ones. These observations are valid for all the tested mocks.

All these comparisons reveal that the best reconstructions (after reconstructions obtained with original, i.e. without error mocks) are obtained with 1.5-Corrected mocks. They enable reconstructions close to the best reconstructions possible at less than 100 \kms\ (2 \hMpc\ in terms of displacement). For completeness, WF full velocity fields obtained with the original mocks are compared with velocity fields of reference simulations. This allows a measurement of the total effectiveness of the method including the WF technique applied to a mock similar to {\it cosmicflows-2}. Table \ref{Tbl:1} gathers the average scatter around the 1:1 linear relation in column 10. A sum in quadrature of the 1-$\sigma$ scatters obtained with a comparison between velocity fields obtained from original mocks / reference simulations and from original / 1.5-Corrected mocks establishes that reconstructed velocity (displacement) fields obtained with 1.5-Corrected mocks are good at 100-150 \kms\ (2-3 \hMpc), the lowest limit that can be reached using linear theories.

\begin{table*}
\begin{center} 
\begin{tabular}{llllllllll}
\hline
\hline
&1 & 2 & 3 & 4 & 5 & 6 & 7 & 8 & 9 \\
Catalog & Ref. Simu & Original & Biased & Corrected & 1.5-Corrected & Original & Biased& Corrected & 1.5-Corrected   \\
 & \kms & \kms & \kms & \kms & \kms & \% & \% & \% & \% \\
\hline
\hline
errorMock &-90 &  -80  & -761 (-780) $\pm$ 80 &  -72 (-60) $\pm$ 37 & -159 (-150) $\pm$ 48 & 99 & 1 & 80 (82) $\pm$ 3 
& 85 (87) $\pm$ 8 \\
haloMock & -90 & -71 (-70) $\pm$ 19 & -816 (-810) $\pm$ 50 & -63 (-60) $\pm$ 26 & -142 (-140) $\pm$ 32 & 97 (98) $\pm$ 2 & 1 & 81 (82) $\pm$ 4 & 88 (88) $\pm$ 8 \\
simuMock & -26 (-80) $\pm$ 159 & -46 (-80) $\pm$ 107 & -804 (-830) $\pm$ 79 & -18 (0) $\pm$ 58 & -60 (-70) $\pm$ 83 & 97 (98) $\pm$ 2 & 1 (1) $\pm$ 1  & 80 (80) $\pm$ 2 & 88 (95) $\pm$ 12 \\
All   & -77 (-90) $\pm$ 70 & -70 (-80) $\pm$ 47 & -792 (-790) $\pm$ 71 & -58 (-60) $\pm$ 42 & -133 (-140) $\pm$ 62 & 98 (99) $\pm$ 2 & 2 (1) $\pm$ 2 & 80 (81) $\pm$ 3 & 87 (87) $\pm$ 9 \\
\hline
\hline
CF2 & / & / & -819 & 71 & -36 &/ & 3 & 77 & 94   \\
\hline
\end{tabular}

\begin{tabular}{l@{ }@{ }@{ }@{ }@{ }@{ }l@{ }@{ }@{ }@{ }@{ }@{ }@{ }@{ }@{ }l@{ }@{ }@{ }@{ }@{ }@{ }@{ }@{ }@{ }@{ }@{ }@{ }l@{ }@{ }@{ }@{ }@{ }@{ }@{ }@{ }@{ }@{ }@{ }@{ }l}
\hline
&10 & 11 & 12 & 13  \\
 & Ref.Simu/ & Original/Biased & Original/Corrected & Original/1.5-Corrected \\
&Original Full & Full / Div \hspace{2.5cm} d   & Full / Div \hspace{2.5cm} d   & Full / Div \hspace{2.5cm} d \\
& \kms  &  \kms / \kms \hspace{0.8cm} unit of density &  \kms / \kms \hspace{0.8cm} unit of density &  \kms / \kms \hspace{0.8cm} unit of density \\
 & \hMpc & \hMpc\ / \hMpc \hspace{2.3cm} & \hMpc\ / \hMpc \hspace{2.3cm} & \hMpc\ / \hMpc \hspace{2.3cm} \\
\hline
\hline
errorMock & 84 &  89 (91) $\pm$ 5 / 69 (70) $\pm$ 3   \hspace{0.8cm} 0.14 &  79 (78) $\pm$ 2 / 81 (80) $\pm$ 2 \hspace{0.8cm} 0.15 & \textbf{71 (70) $\pm$ 3 / 69 (69) $\pm$ 2} \hspace{0.8cm} 0.14    \\
		&	1.7	& 1.8 (1.9) $\pm$ 0.1 / 1.4 (1.4) $\pm$ 0.1  \hspace{0.8cm} & 1.6 (1.6) $\pm$ 0.1 / 1.6 (1.6) $\pm$ 0.1  \hspace{0.8cm} & 1.4 (1.4) $\pm$ 0.1 / 1.4 (1.4) $\pm$ 0.1  \hspace{0.8cm} \\
haloMock  & 86 (87) $\pm$ 2  & 97 (98) $\pm$ 5 / 73 (73) $\pm$ 3 \hspace{0.8cm} 0.14 & 79 (80) $\pm$ 3 / 81 (81) $\pm$ 3 \hspace{0.8cm} 0.15 & \textbf{72 (72) $\pm$ 3 / 71 (71) $\pm$ 4} \hspace{0.8cm} 0.14 \\
		& 1.8 (1.8) $\pm$ 0.1 & 2.0 (2.0) $\pm$ 0.1 / 1.5 (1.5) $\pm$ 0.1 \hspace{0.8cm} & 1.6 (1.6) $\pm$ 0.1 / 1.7 (1.7) $\pm$ 0.1 \hspace{0.8cm} & 1.5 (1.5) $\pm$ 0.1 / 1.4 (1.5) $\pm$ 0.1 \hspace{0.8cm} \\
simuMock  & 85 (85) $\pm$ 1 & 95 (94) $\pm$ 5 / 77 (77) $\pm$ 4 \hspace{0.8cm} 0.14& 83 (85) $\pm$ 5 / 80 (79) $\pm$ 5 \hspace{0.8cm} 0.15 & \textbf{81 (86) $\pm$ 7 / 71 (68) $\pm$ 5} \hspace{0.8cm} 0.14\\ 
 		& 1.8 (1.8) $\pm$ 0.1 & 2.0 (2.0) $\pm$ 0.1 / 1.6 (1.6) $\pm$ 0.1 \hspace{0.8cm} & 1.7 (1.7) $\pm$ 0.1 / 1.6 (1.6) $\pm$ 0.1 \hspace{0.8cm} & 1.6 (1.7) $\pm$ 0.1 / 1.5 (1.4) $\pm$ 0.1 \hspace{0.8cm} \\
All    & 85 (84) $\pm$ 2 & 93 (92) $\pm$ 6 / 72 (72) $\pm$ 4 \hspace{0.8cm} 0.14 & 80 (79) $\pm$ 3 / 81 (80) $\pm$ 3 \hspace{0.8cm} 0.15  & \textbf{73 (72) $\pm$ 6 / 70 (69) $\pm$ 3} \hspace{0.8cm} 0.14 \\
	& 1.7 (1.7) $\pm$ 0.1 & 1.9 (1.9) $\pm$ 0.1 / 1.5 (1.5) $\pm$ 0.1 \hspace{0.8cm} & 1.6 (1.6) $\pm$ 0.1 / 1.6 (1.6) $\pm$ 0.1 \hspace{0.8cm} & 1.5 (1.5) $\pm$ 0.1 / 1.4 (1.4) $\pm$ 0.1 \hspace{0.8cm} \\
\hline
\hline
\end{tabular}
\end{center}
\caption{mean (median) $\pm$ standard deviation of monopole terms, confidence levels obtained with F-Tests for Gaussianity and cell-to-cell 1-$\sigma$ scatters computed for the Wiener-Filter reconstructed overdensity, full and divergent velocity fields obtained with biased, corrected and original mocks. The first column identify the mock series: mocks in the errorMock section differ only on the error realization (only one original mock corresponds to that series) ; mocks belonging to the haloMock category have a different halo selection (only one reference simulation corresponds to that series) ; mocks in simuMock are built from different constrained simulations. 'All' gathers the means, medians and standard deviations over all the mocks. 'CF2' gives values for the observational catalog, {\it cosmicflows-2}. (1) monopole term at 150 \hMpc\ of the velocity fields of reference simulations, \kms, (2) to (5) monopole term value at 150 \hMpc\ of  reconstructed velocity fields obtained with the Wiener-Filter applied to original, biased, corrected and 1.5-Corrected mocks, \kms, (6) to (9) confidence level obtained with F-Tests for Gaussianity applied to the original, biased, corrected and 1.5-Corrected mock catalog. A Gaussian with a variance of 300 \kms is used for the tests. The higher the confidence level the better the agreement between a Gaussian with a variance of 300 \kms and the distribution of radial peculiar velocities, \% (10) 1-$\sigma$ scatter of cell-to-cell comparisons in 320 \hMpc\ between full velocity fields of the reference simulations and obtained with the Wiener-Filter technique applied to original mocks, \kms\ (first line) - \hMpc\ (second line), (11) to (13) 1-$\sigma$ scatter of cell-to-cell comparisons in 320 \hMpc\ between Wiener-Filter reconstructed full, divergent velocity and overdensity fields obtained with original and biased, corrected, 1.5-Corrected mocks, \kms\ and unit of overdensity (first line) - \hMpc\ (second line). Bold figures represent reconstructions the closest to reconstructions obtained with original mocks. These are reconstructions obtained with the 1.5-Corrected mocks.}
\label{Tbl:1}
\end{table*}

\section{Application to cosmicflows-2 catalog}

\begin{figure*}
\includegraphics[scale=1]{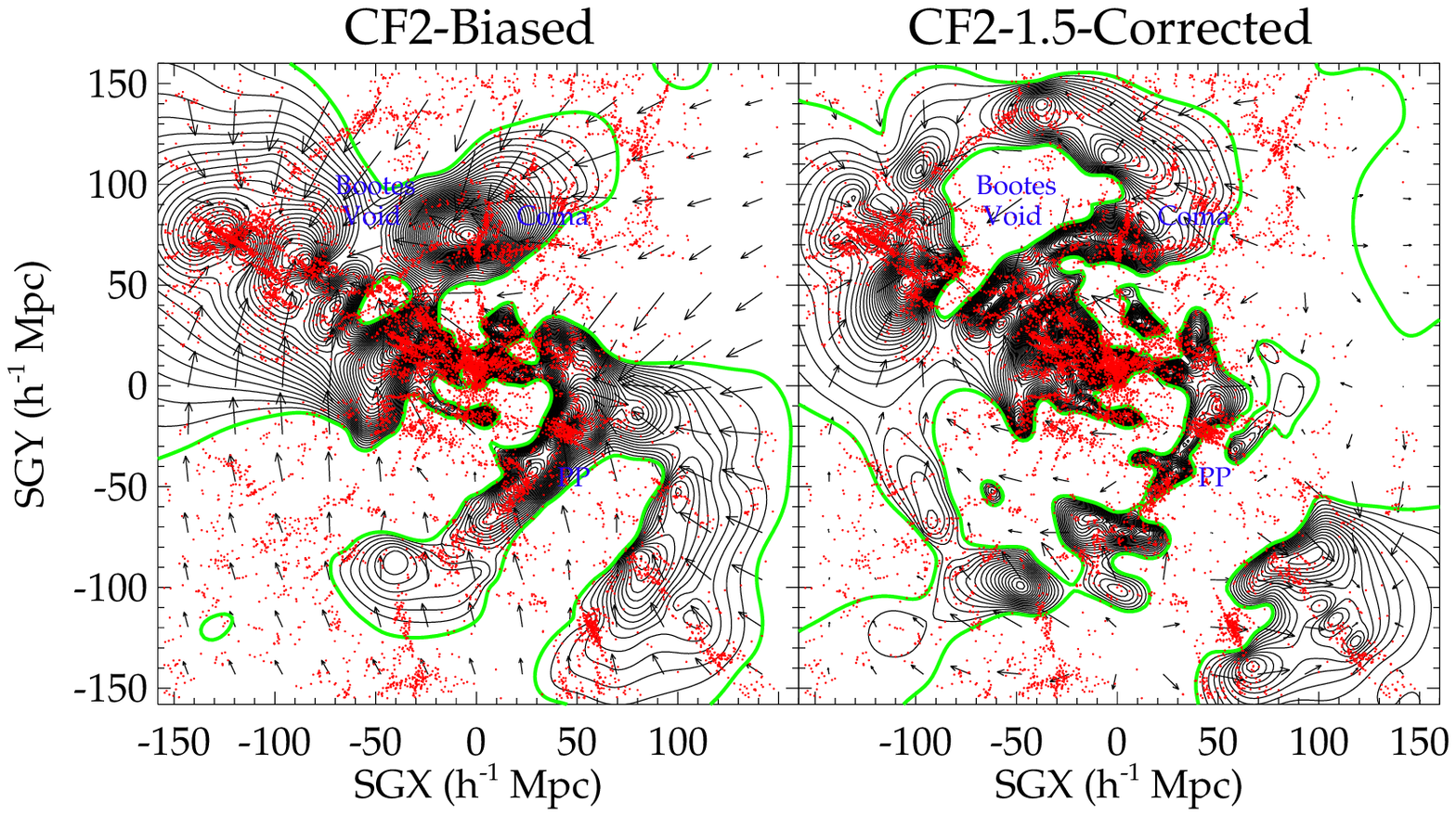}
\includegraphics[scale=1]{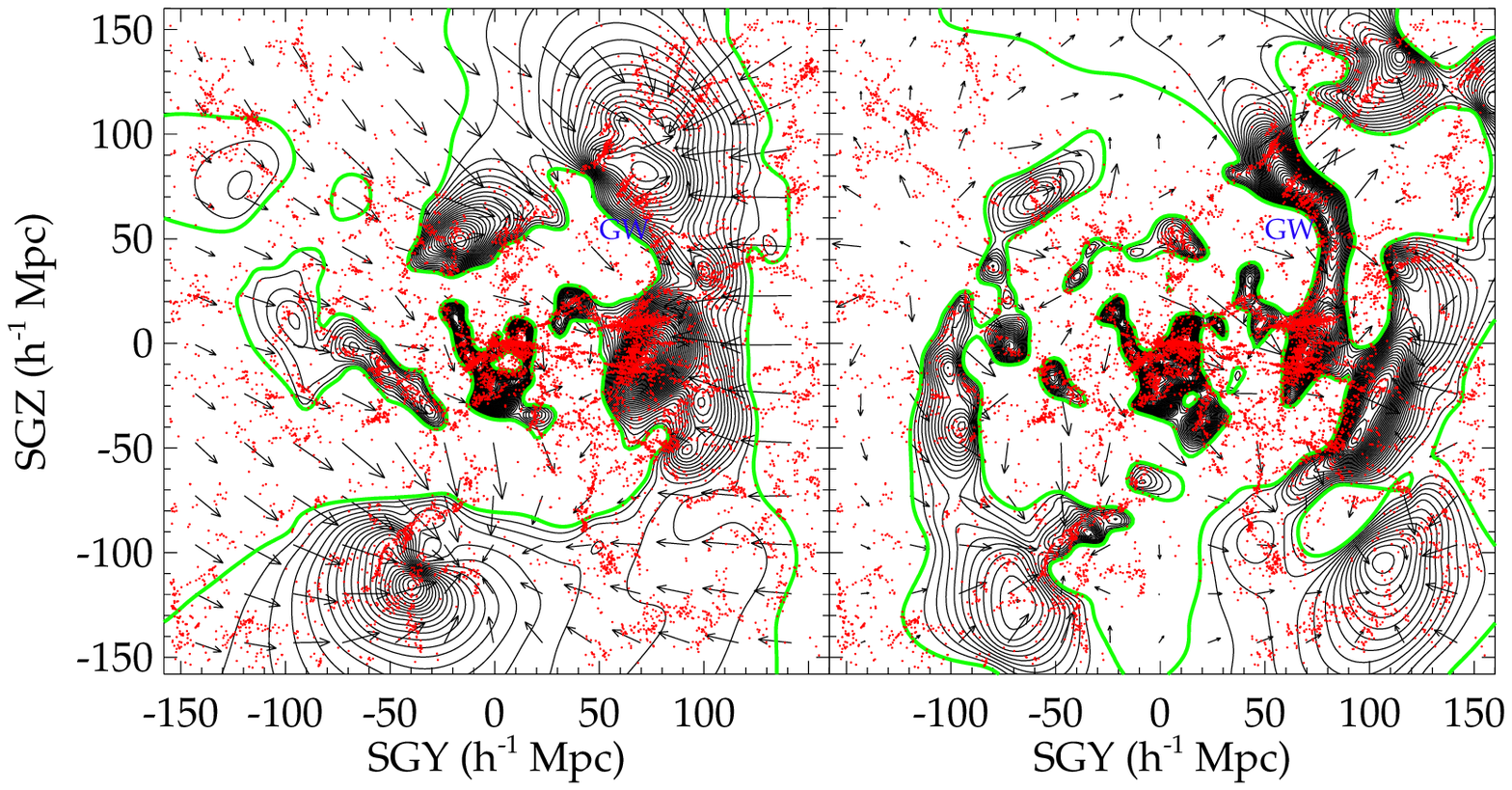}
\caption{Supergalactic XY (top) and YZ (bottom) plane of the reconstructed velocity (arrows) and overdensity (contours) fields obtained with the raw or biased {\it cosmicflows-2} catalog (left) and the 1.5-corrected {\it cosmicflows-2} catalog (right). The green color stands for the mean density. Red dots are galaxies in a 5 \hMpc\ thick slice from the 2MASS redshift catalog. They are superimposed to the reconstructions for comparison purposes solely. A few structures are identified such as Coma, Perseus Pisces (PP), Bootes Void and the Great Wall (GW). For instance, the contour of Bootes Void are better defined and the Great Wall matches more galaxies from the redshift catalog in the reconstruction obtained with the corrected catalog.}
\label{WFCF2}
\end{figure*}

To perfect the tests, we apply the correction, then the WF, to {\it cosmicflows-2}, and evaluate the monopole terms of the different reconstructed velocity fields. Figure \ref{symasym} shows that the correction enables us to recover a close to a 
Gaussian radial peculiar velocity distribution for {\it cosmicflows-2}. Actually, a F-test confirms that a Gaussian model of variance 300 \kms\ cannot be rejected at more than the 6 \% confidence level while it can be rejected at more than the 95\% confidence level before any correction. The CF2 line in Table \ref{Tbl:1} gives the monopole term values which are decreased by a factor larger than 10 for the corrected catalog when compared to CF2-biased. Figure \ref{WFCF2} shows the supergalactic XY and YZ plane of the reconstructed velocity (arrows) and overdensity (contours) of the Local Universe obtained with CF2-Biased and CF2-1.5-Corrected. Galaxies in a 5 \hMpc\ thick slice (red dots) from the 2MASS redshift catalog \citep{2012ApJS..199...26H} are superimposed to the reconstructions (solely for comparison purposes). First the observed infall in the reconstruction obtained with the CF2-biased catalog is not present anymore in the reconstruction obtained with the corrected catalog in agreement with monopole values. Second structures are more sharply defined. To guide the reader's eye, some of them are identified with blue names. For instance one can notice that the Coma region is more condensed or that the Great Wall (GW) is in better agreement with galaxies from the redshift survey. The same is also true with voids such as Bootes void which has more well-defined contours, contours which follow the galaxy distribution.

Observing these changes, one can wonder whether the Laniakea supercluster of galaxies discovered by \citet{2014Natur.513...71T} is affected. Before defining Laniakea, it can be noted that in their paper, \citet{2014Natur.513...71T} apply a Malmquist Bias correction to reduce the infall in the reconstruction obtained with the {\it cosmicflows-2} catalog. They show that their correction affects only the global infall and that structures are unaffected, i.e. Laniakea is not the result of the correction. Their correction consists in changing only distances and keeping radial peculiar velocities unchanged. With this paper, we propose to change velocities to remove also the asymmetry bias. The Laniakea supercluster of galaxies is defined as a local basin of attraction and every flow which converges towards it, i.e. the boundary of the Laniakea supercluster corresponds to a zero velocity isosurface. Galaxies within that zone fall onto the local attractor while galaxies outside of that zone are going in opposite direction. This surface is derived with the divergent part of the velocity field, i.e. with densities solely in a 100 \hMpc\ radius sphere centered around [-47,13,-5] \hMpc. We have shown with cell-to-cell comparisons that biases affect mostly the full reconstructed velocity field and that the divergent reconstructed velocity field is relatively unaffected by the biases. The correction technique proposed in this paper reduces drastically the average 1-$\sigma$ scatter in the cell-to-cell comparison of full velocity fields while the average 1-$\sigma$ scatter in the cell-to-cell comparison of the divergent parts of the velocity fields is mostly unchanged. The result is that the Laniakea supercluster of galaxies is globally unchanged. We compute the divergent velocity field within a 100 \hMpc\ radius sphere around [-47,13,-5] \hMpc\ and find back the Laniakea supercluster of galaxies as expected. 
 
\section{Summary \& Conclusion}

The Local Universe is composed of a multitude of structures (nodes, sheets, filaments, voids) which need to be studied to better understand how gravitational motions induce the formation of such a complex environment. In that respect, galaxy distances and derived radial peculiar velocity catalogs constitute a valuable source of information to map the Local Universe. The Cosmicflows project develops such catalogs. These catalogs, however, suffer from biases which, when not corrected for, results in spurious velocity infall onto the Local Volume. \citet{2013AJ....146...86T} themselves warn us that these issues must be addressed if the catalog is used to infer matter and velocity distributions. These biases are due to the inhomogeneity of the Local Universe and to the fact that distances in megaparsecs are derived from distance moduli in magnitudes via a logarithmic function. As a result, although errors on distance moduli in magnitudes have a symmetric distribution, the distribution of uncertainties on distance estimate in megaparsecs is asymmetric. Consequently, the radial peculiar velocity distribution has a larger negative tail than a positive one. This asymmetric bias, in addition to Malmquist effects, inserted in the Bayesian Wiener-Filter reconstruction technique, results in a spurious infall onto the Local Volume.

This paper proposes a correction to minimize the effect of biases observed in the reconstructed fields obtained with radial peculiar velocity catalogs such as the second catalog of the Cosmicflows project. The approach is based on the radial peculiar velocity distribution rather than on the radial distance distribution. Distance estimates are thus overall corrected retroactively after peculiar velocities are. As radial peculiar velocity distributions have been shown to be Gaussians, an iterative method is proposed to reduce spurious non-Gaussianities thus correct radial peculiar velocities and as a result distances. The correction is developed and tested on a suite of mocks. These mocks are built out of constrained cosmological simulations which resemble the Local Universe to limit the cosmic variance. They are made to mimic as closely as possible the error, data and uncertainty distributions of the second catalog of the Cosmicflows project, so as to reproduce as realistically as possible biases' effects affecting radial peculiar velocity catalogs of our neighborhood. 

Before the addition of errors, Gaussian distributions are found in all the mock catalogs built out of reference simulations. The Gaussian distributions are similar in terms of variance and amplitude. As a matter of fact, F-tests revealed that a theoretical Gaussian of variance 300 \kms\ is in agreement with radial peculiar velocity distributions of every original (i.e. without errors) mock at the average 98 $\pm$ 2 \% confidence level. To correct a radial peculiar velocity from a biased mock, both its probability to belong to this theoretical Gaussian and its (weighted) uncertainty are taken into account. Corrected radial peculiar velocities have distributions in agreement with the theoretical Gaussian at the average 80 $\pm$ 3 \% confidence level according to F-tests. In comparison, such a Gaussian model for non-corrected (i.e. biased) radial peculiar velocity distributions can be rejected at an average confidence level of 98 $\pm$ 2 \%. Corrected radial peculiar velocities allow the derivation of new distances, required for consistency and to take care of the Malmquist biases. The residuals of these corrected distances with respect to the true distances have standard deviations of approximately 4 \hMpc, compared to a $\sim$ 15 \hMpc\ before correction. 

The Wiener-Filter reconstruction technique was applied to the suite of mocks. Cell-to-cel comparisons between WF reconstructions obtained with corrected and original (in the sense that no error was added) mocks reveal that scatters are decreased by 10 - 20 \kms\ in comparison with cell-to-cell scatters obtained for WF reconstructions obtained with biased and original mocks, however, corrected peculiar velocities are not high enough (in absolute value) to overpower the Wiener-Filter smoothing. Tilts were observed in every cell-to-cell comparison between reconstructed velocity fields obtained with original and corrected mocks. These tilts are due to the fact that reconstructed velocities are smaller (in absolute value) than true velocities by a factor $\sim$ 1.5. Tests revealed that multiplying reconstructed velocity fields - obtained with corrected mocks - by 1.5 removed these tilts. Still Wiener-Filter reconstructions abide by a prior. Accordingly, corrected radial peculiar velocities are multiplied by 1.5 \emph{before} being inserted in the Wiener-Filter together with their corresponding distances. The Gaussian theoretical model is still valid at the average 87 $\pm$ 9 \% confidence level according to F-tests. A 5\% fractional error in agreement with the upper limit of the median error values found for the different built mock catalogs, when comparing true and corrected distances, is attributed to datapoint. The Wiener-Filter technique is applied to these new corrected mocks. Scatters obtained with cell-to-cell comparisons are decreased by an additional 5 \kms\ and tilts are no longer evident.

The series of built mocks highlights the independence of the method outcome on the added error realization, on the datapoint selection and on the constrained reference simulation. With constrained simulations, resembling the Local Universe, and associated mocks, mimicking {\it cosmicflows-2}, the method robustness and accuracy have been tested.  The resulting reconstructed overdensity and velocity fields are very close to those obtained when applying the Wiener-Filter to original (i.e. without error) mocks (at $<$ 100 \kms\ or 2 \hMpc) and to velocity fields of reference constrained cosmological simulations (at $<$ 150 \kms\ or 3 \hMpc).  After correction, the infall is drastically reduced as monopole terms are divided by 5 - 6. These monopole terms are related to the flow into the Local Volume, a large positive value or conversely a large negative value implies a large outflow, infall respectively.  Structures are more sharply defined. Applied to the observational {\it cosmicflows-2} catalog, the method confirms its ability to decrease the spurious infall onto the Local volume by dividing the monopole term of the reconstructed velocity field by a factor $\sim$ 10. The corrected {\it cosmicflows-2} catalog distribution is in agreement with the theoretical Gaussian at the 94 \% confidence level. Structures like the Great Wall and Bootes Void contours are better defined. These changes motivate the examination of whether the Laniakea supercluster of galaxies defined as a local basin of attraction and flows converging onto it and discovered by \citet{2014Natur.513...71T}, who chose to modify distances, is affected by the correction proposed in this paper. Because the divergent velocity field, due to densities solely in a volume, is quite unaffected by the biases and because the correction technique proposed in this paper changes only the full velocity fields (divergent velocity fields are quite stable), the Laniakea Supercluster of galaxies, derived with the divergent velocity field computed within a 100 \hMpc\ radius sphere centered on [-47,13,-5] \hMpc, survives the correction proposed in this paper. 

The correction method applied to the second catalog of the Cosmicflows project result in a dataset of sufficient quality to build constrained initial conditions, based on the process described in \citet{2014MNRAS.437.3586S}, of similar quality from it. Simulations of the Local Universe constrained by {\it cosmicflows-2} will be released in a subsequent paper.

\section*{Acknowledgements}
The referee, Michael Strauss, provided useful comments which contributed to improve the paper. Fruitful discussions with Yehuda Hoffman, Brent Tully, Helene Courtois and Stefan Gottl\"ober were appreciated. Peter Creasey made helpful remarks and provided a careful reading of the manuscript. This research was supported by the Alexander von Humboldt Foundation. The simulations were performed at the Leibniz Rechenzentrum (LRZ) in Munich.


\bibliographystyle{mnras}

\bibliography{biblicomplete}
\section*{Appendix}

Let us consider a particle at Lagrangian coordinates $\mathbf{x}_L$ and the coordinates of the grid point occupied by the particle at a time $t=0$, $\mathbf{x}_E$ or Eulerian coordinates. At a later time $t$, the particle is located on the grid at $\mathbf{x}_E(t)=\mathbf{x}_L(\mathbf{x}_E)+\mathbf{\psi}(\mathbf{x}_E,t)$ where $\mathbf{\psi}(\mathbf{x}_E,t)$ is the displacement field from the initial position. The Zel'dovich approximation \citep{1970A&A.....5...84Z} stipulates that this displacement field can be approximated by:
\begin{equation}
 \mathbf{\psi}(\mathbf{x}_E,t)= D_+(t)\mathbf{\psi}_0(\mathbf{x}_E)
 \label{eqCh2:ZelApp}
 \end{equation}
where $\mathbf{\psi}_0(\mathbf{x}_E)$ is the initial displacement field. In other words, the displacement field behaves similarly to the perturbation density field. Its direction is frozen and it grows with time. 

Because the peculiar velocity field $\mathbf{v}$ is the time derivative of the distance, $D(t)=a(t)\mathbf{x}_E(t)$, minus expansion, $\dot a \mathbf{x}_E(t) = \dot a/a D(t)$:
\begin{equation}
\mathbf{v}(\mathbf{x},t)=a(t) \mathbf{\dot x}_E(t) = a \dot \psi(\mathbf{x}_E,t) = a \times \frac{\mathbf{\psi}(\mathbf{x_E},t)}{D_+(t)} \frac{d D_+}{d t} \times \frac{d t}{d a} \dot a = \dot a f  \mathbf{\psi}(\mathbf{x_E},t)
\label{eqCh2:VPecIniti}
\end{equation}
where $f=\frac{d(\mathrm{ln}D_+)}{d(\mathrm{ln}a)}$ is the growth rate, the displacement field can be reached through the peculiar velocity field. The 'dot' notation corresponds to the time derivative.

Today, $\dot a$ is $H_0$. Consequently, in \hMpc, the conversion factor between the velocity field and the displacement field is 100$f$ where f is the growth rate.
\end{document}